\documentclass[10pt]{article}

\usepackage[numbers,square]{natbib}

\usepackage{xcolor}
\usepackage{epsfig}
\usepackage{cancel}
\usepackage{caption}
\usepackage{feynmf} 
\usepackage{amssymb}
\usepackage{amsfonts}
\usepackage{epsf}
\usepackage{rotating}
\usepackage{graphicx}
\usepackage{amsmath}
\usepackage{fancyhdr}
\usepackage{subfigure}
\usepackage{graphics}
\usepackage{pstricks}
\usepackage{color}
\usepackage{frontespizio}
\usepackage{hyperref}
\hypersetup{
	colorlinks,
	citecolor=green,
	filecolor=black,
	linkcolor=blue,
	urlcolor=black
}
\usepackage{type1ec}
\usepackage[T1]{fontenc}
\usepackage{lettrine}
\usepackage{bbold}
\usepackage{calligra}
\usepackage{tikz}
\usepackage{subfigure}
\usepackage{mathrsfs}
\usepackage{curve2e}
\usepackage{setspace}
\usepackage{indentfirst}
\usepackage{emptypage}
\usepackage[babel]{csquotes} 
\usepackage[font=small,labelfont=bf,labelsep=quad]{caption} 
\usepackage{graphicx} 
\usepackage{listings} 
\usetikzlibrary{patterns}
\usepackage{relsize}
\usetikzlibrary{intersections,positioning}
\usetikzlibrary{decorations.pathmorphing,decorations.markings,arrows,positioning}
\usepackage{braket}
\usepackage{mathrsfs}
\usepackage{stackengine}
\usepackage{calc}
\newlength\shlength
\newcommand\xshlongvec[2][0]{\setlength\shlength{#1pt}%
	\stackengine{-5.6pt}{$#2$}{\smash{$\kern\shlength%
			\stackengine{7.55pt}{$\mathchar"017E$}%
			{\rule{\widthof{$#2$}}{.57pt}\kern.4pt}{O}{r}{F}{F}{L}\kern-\shlength$}}%
	{O}{c}{F}{T}{S}}

\newcommand*{\Scale}[2][4]{\scalebox{#1}{$#2$}}%
\newcommand{\app}[4]{F_{\!#1}\!
  \left(\left.\substack{\Scale[1]{ #2} \\[1.5ex] \Scale[1]{#3}}\right| #4 \right) }

\newcommand{\hpg}[5]{{}_{#1}\mbox{\rm F}_{\!#2}\!
  \left(\left.\substack{\Scale[1]{ #3} \\[1.5ex] \Scale[1]{#4}}\right| #5 \right) }

\newcommand{\hpgo}[2]{{}_{#1}\mbox{\rm F}_{\!#2}}

\newcommand{\equal}{&\!\!=\!\! &}
\IfFileExists{dsfont.sty}
{\usepackage{dsfont}
	\let\mathbb=\mathds
	\newcommand{\id}{\mathds{1}}}
{\typeout{Package dsfont.sty was not found, using alternative macros.}
	\let\mathds=\mathbb
	\newcommand{\id}{\mbox{1 \kern-.59em {\rm l}}}}

\usepackage{slashed}
\usepackage{units}
\usepackage{latexsym}
\usepackage{setspace}
\renewcommand{\Re}{\textrm{Re}}
\renewcommand{\Im}{\textrm{Im}}
\newcommand{\nn}{\nonumber}
\let\a=\alpha   \let\b=\beta   \let\g=\gamma   \let\d=\delta
         
    \let\k=\kappa  \let\l=\lambda  \let\m=\mu
\let\n=\nu                 
\let\s=\sigma        
         
\let\D=\Delta
\let\d=\delta
\let\s=\sigma
\renewcommand{\a}{\alpha}

\newcommand{\ul}{\underline}
\def\nbox#1#2{\vcenter{\hrule \hbox{\vrule height#2in
			\kern#1in \vrule} \hrule}}
\def\sq{\,\raise.5pt\hbox{$\nbox{.09}{.09}$}\,}
\def\sqb{\,\raise.5pt\hbox{$\overline{\nbox{.09}{.09}}$}\,}

\newcommand{\bea}{\begin{eqnarray}}
\newcommand{\eea}{\end{eqnarray}}
\newcommand{\be}{\begin{equation}}
\newcommand{\ee}{\end{equation}}

\newcommand{\bes}{\begin{subequations}}
	\newcommand{\ees}{\end{subequations}}

\def\nn{\nonumber\\}

\numberwithin{equation}{section}

\usepackage{accents}

\makeatletter
\newcommand{\xLine}[2][]{\ext@arrow 0359\Rightarrowfill@{#1}{#2}}
\makeatother
\xdefinecolor{darkgreen}{RGB}{102, 204, 70}
\xdefinecolor{darkblue}{RGB}{0, 0, 153}

\begin{document}

\begin{center}

{\bf \Large The Generalized Hypergeometric Structure of the Ward Identities
of CFT's in Momentum Space in $d  > 2$\\ }
\vspace{2.5cm}
{\bf Claudio Corian\`{o} and Matteo Maria Maglio \\ }
\vspace{0.5cm}
{\it Dipartimento di Matematica e Fisica "Ennio De Giorgi",\\ 
Universit\`{a} del Salento and INFN-Lecce,\\ Via Arnesano, 73100 Lecce, Italy\footnote{claudio.coriano@le.infn.it, matteomaria.maglio@le.infn.it}
}
\end{center}
\begin{abstract}
We review the emergence of hypergeometric structures (of $F_4$ Appell functions) from the conformal Ward identities (CWIs) in conformal field theories (CFTs) in dimensions $d > 2$. We illustrate the case of scalar 3- and 4-point functions. 3-point functions are associated to hypergeometric systems with 4 independent solutions. For symmetric correlators they can be expressed in terms 
of  a single 3K integral - functions of quadratic ratios of momenta - which is a parametric integral of three modified Bessel $K$ functions. In the case of scalar 4-point functions, by requiring the correlator to be conformal invariant in coordinate space as well as in some dual variables (i.e., dual conformal invariant), its explicit expression is also given by a 3K integral, or as a linear combination of Appell functions which are now quartic ratios of momenta. Similar expressions have been obtained in the past in the computation of an infinite class of planar ladder (Feynman) diagrams in perturbation theory, which, however, do not share the same (dual conformal/conformal) symmetry of our solutions.  We then discuss some hypergeometric functions of 3 variables, which define 8 particular solutions of the CWIs and correspond to Lauricella functions. They can also be combined in terms of 4K integral and appear in an asymptotic description of the scalar 4-point function, in special kinematical limits. 

\end{abstract}

\vspace{3cm}
Invited contribution to appear in {\em Axioms}, {\em "Geometric Analysis and Mathematical Physics"}\\ \hspace{1cm} (Ed. Sorin Dragomir)

\newpage
\section{Introduction} 
Conformal symmetry has been important in the study of critical phenomena as well as in string theory, over about fifty years \cite{Kastrup:2008jn,DiFrancesco:1997nk}. In ordinary string theory it has played a key role in its two-dimensional version $(d=2)$, with the identification of an infinite dimensional Virasoro algebra which has been crucial for the characterization of the dynamics of the theory \cite{Belavin:1984vu}.\\
The interest in conformal field theories (CFTs), however, has developed in parallel but also independently of string theory, and in $d=2$ the presence of such enhanced symmetry has allowed, on the other hand, to come up with important predictions about the behaviour of several statistical models at their critical points, accounting for many of their universality properties. This, in general, allows the computations of their critical exponents and the characterization of their correlation functions \cite{Polyakov:1970xd,Henkel:1999qx}. \\
CFT descriptions, as just mentioned, describe systems which develop, at a critical point, long range quantum correlations with power-like decay laws as functions of their separation (coordinate) points. The absence of any dimensionful parameter in the theory only allows correlators characterised by an algebraic - rather than exponential - decay as function of  distance.
These are controlled by some exponents related to the scaling dimensions of the field operators appearing in the quantum average. A set of primary fields - together with their descendents - and their correlation functions, provide information about the quantum fluctuations of the theory. 
The products of two primary operators describing such fluctuations obey some operatorial relations defined by an operator product expansion (OPE) which is entirely dependent on the parameter of a specifc CFT (couplings, scaling dimensions), also called "the conformal data" \cite{Dolan:2000ut}.   \\
The OPE is at the basis of the so-called "conformal bootstrap" program \cite{Ferrara:1973yt,Dolan:2000ut,Poland:2018epd,Poland:2016chs,Simmons-Duffin:2016gjk}, whose goal is to generate correlators of higher point starting from the lower point ones (2 and 3-point functions). 2- and 3- point functions are essentially fixed by the symmetry, which acts on their explicit expressions via a set of Conformal Ward identities (CWIs) \cite{Osborn:1993cr,Erdmenger:1996yc}. These can be generalized to $n$-point functions \cite{Irges:2020lgp}. In coordinate space - we assume them to be defined in $\mathbb{R}^d$ - such equations are of first order and become of second order in momentum space.\\
The goal of this review is to describe the CWIs of 3-point functions in momentum space, illustrating the hypergeometric character of such equations, which could be of interest from the mathematical side.

Traditionally, the critical behaviour of a certain theory has been investigated using the renormalization group approach \cite{Wilson:1973jj,Wilson:1993dy}. In this approach, starting from a certain Hamiltonian of a given system, one builds a sequence of Hamiltonians, each defined at a certain distance scale $(\lambda_n)$, $H_n\equiv H_{\lambda_n}$, through the process of rescaling and decimation of its degrees of freedom, in order to describe the flow of the theory as we vary the fundamental scale. One looks for the fixed points of the sequence $H_{\bar{n}+1}=H_{\bar{n}}$ (for all $n>\bar{n}$), with $H_{\bar{n}}$ the fixed-point Hamiltonian. The scaling dimensions of the theory are identified by an analysis of its quantum fluctuations using the fixed-point Hamiltonian of the model. \\
Conformal symmetry defines an independent path compared to the previous one. 
Exploiting the fact that at certain critical point a given system is characterised by a dynamics which is controlled by such symmetry, one is able to describe the behaviour of its correlation functions without any additional input.  By a use of the OPE in a CFT and its constraints in various channels - there are three channels for a 4-point function, for instance - it is possible to derive  - independently of any renormalization group analysis - the critical exponents of the theory. This approach requires full knowledge of the conformal blocks (or conformal partial waves) of a certain theory, which is a topic of central relevance in the study of any CFT \cite{Poland:2018epd}.

\subsection{The momentum space analysis}
Most of this analysis, so far, has been developed in coordinate space.\\
One may wonder why one should bother to reformulate such CFTs in momentum space or in other spaces, such as Mellin space \cite{Gopakumar:2016wkt,Sleight:2019mgd,Sleight:2019hfp}. These new approaches are currently under investigation from many different sides \cite{Gillioz:2019lgs,Gillioz:2019iye,Isono:2018rrb,Isono:2019wex,Albayrak:2018tam,Bautista:2019qxj,Albayrak:2019yve,Albayrak:2020isk}, including their direct links to cosmology\cite{Arkani-Hamed:2018kmz,Baumann:2019oyu,Arkani-Hamed:2017fdk,Benincasa:2018ssx,Arkani-Hamed:2018bjr}.

The reason is twofold. First, CFT correlators in momentum space offers a description of a correlation function which is quite close to that provided by ordinary quantum field theories, in terms of scattering amplitudes and of $S$-matrix elements, in which conformal symmetry plays a significant guiding role \cite{Broadhurst:1993ib,Usyukina:1992jd,Broadhurst:1993ru}. The second is related to issues concerning the UV behaviour of such theories, described when all the points of a given correlator coalesce. This induces a breaking of the classical conformal symmetry at quantum level, with the appearance of a conformal anomaly \cite{Capper:1975ig,Deser:1976yx,Riegert:1984kt}. 

The interest in CFT in higher dimensions has grown significantly after the formulation of the Anti De Sitter (AdS) CFT duality \cite{Maldacena_1999,Anand:2019lkt,Albayrak:2018tam}.
For $d > 2$, conformal symmetry is finite dimensional, and the dynamics of such CFTs is far less constrained. Nevertheless, the correlation functions of CFTs are constrained by a finite set of conformal Ward identities (CWIs) that we are going to discuss in the next sections. \\
In the case of tensor correlators, additional symmetries induce additional WI's, the canonical WIs, due to Noether symmetries which must also be respected. They are related to the Poincar\'e symmetry.\\
These are hierarchical, and connect $n$-point functions to $n-1$-point functions, and so on.  In this brief review, we are going to illustrate the key steps that take to the identification of a generalized hypergeometric structure which emerges from the equations associated to such CWIs, once we turn to momentum space.\\
We will be focusing our attention and summarize the content of some original work on the subject for scalar \cite{Coriano:2013jba} and tensor 3-point functions in $d=4$ \cite{Bzowski:2013sza,Coriano:2018bsy,Coriano:2018bbe}, which is relevant for the analysis of the implications of conformal symmetry in several field theory contexts. Details and derivations can be found in those works. Our goal will simply be to outline some of the main results of these analysis which may raise the interest of mathematicians.\\ 
Discussions of hypergeometric systems of two variables can be found, for instance, in \cite{Vidunas1}. A more extensive review of these developments, with a detailed description of the results that we are going to summarize here, will be presented by us elsewhere.\\

\section{Conformal Ward Identities (CWIs)}
 For a discussion of the general features of CFTs in d $\geq 2$, we refer to the several reviews which have been published in the last few years \cite{Simmons-Duffin:2016gjk, Penedones:2016voo, Rychkov:2016iqz}. Most of them deal with the analysis of such theories in coordinate space. The momentum space approach to CFT is a more recent area of research. In $d=4$ it has been investigated in \cite{Coriano:2013jba,Bzowski:2013sza} and \cite{Coriano:2018bbe,Coriano:2018bsy, Bzowski:2015pba,Bzowski:2018fql} and in more recent work in \cite{Bzowski:2019kwd, Maglio:2019grh}. The hypergeometric structure of the CWIs has been identified independently in \cite{Coriano:2013jba} and \cite{Bzowski:2013sza}, as already mentioned, in the case of 3-point functions. The identification of generalized hypergeometric solutions of the CWIs for 4-point functions, which share a structure typical of 3-point functions, and of the homogenous solutions of Lauricella type, have been discussed in \cite{Maglio:2019grh}. \\
  The CWIs are composed of special conformal and dilatation WIs, beside the ordinary (canonical) WI's corresponding to Lorentz and translational symmetries, that we are going to specify below. We recall, for instance, that in $d=4$ conformal symmetry is realized by the action of $15$ generators, $10$ of them corresponding to the Poincar\'e subgroup, $4$ to the special conformal transformations and $1$ to the dilatation operator. 
 In the infinitesimal form, they are given by 
\begin{equation}
x'_\mu(x)=x_\mu+a_\mu+\omega_{\mu\nu}x^\nu+\sigma x_\mu+b_\mu x^2-2b\cdot x\,x_\mu, \label{transf}
\end{equation}
and they can be expressed as a local rotation 
\begin{equation}
\frac{\partial x '^{\mu}}{\partial x^\nu}=\Omega(x) R^\mu_{\nu}(x),
\end{equation}
where $\mu=1,2\ldots d$, and $\Omega(x)$ and $R^\mu_\nu(x)$ are, respectively, finite position-dependent rescalings and rotations with
\begin{equation}
\Omega(x)=1-\l(x),\quad \l(x)=\s-2b\cdot x,\label{Om}
\end{equation}
and $b_\mu$ is a constant $d$-vector.
The transformation in \eqref{transf} is composed of the parameters $a_\mu$ for the translations, $\omega_{\mu\nu}=-\omega_{\nu\mu}$ for boosts and rotations, $\sigma$ for the dilatations and $b_\mu$ for the special conformal transformations. The first three terms in \eqref{transf} define the Poincar\'e subgroup, obtained for $\Omega(x)=1$, which leaves invariant the infinitesimal length. For a general $d$, the counting of the parameters of the transformation is straightforward. 
We have $d(d-1)/2$ ordinary rotations associated to a $SO(d)$ symmetry in 
$\mathbb{R}^d$ - with parameters $\omega_{\mu\nu}$ - $d$ translations ($P_\mu$) with parameters $a_\mu$, $d$ special conformal transformations $K^\mu$ (with parameters $b_\mu$), and one dilatation $D$ whose corresponding parameter is $\sigma$, for a total of $(d+1)(d+2)/2$ parameters. This is exactly the number of parameters appearing in general of $SO(2,d)$ transformation. 
Indeed one can embed the actions of the conformal group of $d$ dimensions into a larger $\mathbb{R}^{d+2}$ space, where the action of the generators is linear on the coordinates $x^M$ ($M=1,2,\ldots,d+2$) of such space, using a projective representation. This is at the basis of the so-called embedding formalism. We refer to \cite{Simmons-Duffin:2016gjk} for more details.
By including the inversion $(\mathcal{I})$
\begin{equation}
x_\mu\to x'_\mu=\frac{x_\mu}{x^2},\qquad \Omega(x)=x^2,
\end{equation}
we can enlarge the conformal group to $O(2,d)$. Special conformal transformations can be realized by considering a translation preceded and followed by an inversion. 
\\
We will focus our discussion mostly on scalar primary operators of a quantum CFT, acting on an certain Hilbert space, which under a conformal transformation will transform as 
\begin{equation}
O_i(\mathbf{x})\to O'(\mathbf{x}')=\lambda^{-\Delta_i} O(\mathbf{x})
\end{equation}
with specific scaling dimensions $\Delta_i$.
 We start this excursus on the implication of such symmetry on the quantum correlation functions of a CFT, by considering the simple case of a correlator of $n$ primary scalar fields $O_i(\mathbf{x}_i)$, each of scaling dimension $\Delta_i$
\begin{equation}
\label{defop}
\Phi(\mathbf{x}_1,\mathbf{x}_2,\ldots,\mathbf{x}_n)=\braket{O_1(\mathbf{x}_1)O_2(\mathbf{x}_2)\ldots O_n(\mathbf{x}_n)}.
\end{equation}
In all the equations, covariant variables will be shown in boldface.\\
 3- and 4-point functions (beside 2-point functions) in any CFT are significantly constrained in their general structures due to such CWI's.  For scalar correlators the special CWI's are given by first order differential equations 
\begin{equation}
\label{SCWI0}
K^\kappa(x_i) \Phi(\mathbf{x}_1,\mathbf{x}_2,\ldots,\mathbf{x}_n) =0,
\end{equation}
with 
\begin{equation}
\label{transf1}
K^\kappa(x_i) \equiv \sum_{j=1}^{n} \left(2 \Delta_j x_j^\kappa- x_j^2\frac{\partial}{\partial x_j^\kappa}+ 2 x_j^\kappa x_j^\alpha \frac{\partial}
{\partial x_j^\alpha} \right)
\end{equation}
being the expression of the special conformal generator in coordinate space. \\
The corresponding dilatation WI on the same $n$-point function $\Phi$ is given by 
\begin{equation}
\label{scale12}
D(x_i)
\Phi(\mathbf{x}_1,\ldots \mathbf{x}_n)=0,
\end{equation}
with 
\begin{equation}
\label{scale11}
D(x_i)\equiv\sum_{i=1}^n\left( x_i^\alpha \frac{\partial}{\partial x_i^\alpha} +\Delta_i\right)
\end{equation}
for scale covariant correlators. In the case of scale invariance the dilatation WI takes the form 
\begin{equation}
D_0(x_i)
\Phi(\mathbf{x}_1,\ldots \mathbf{x}_n)=0,
\end{equation}
with $D_0(x_i)$ given by
\begin{equation}
D_0(x_i)\equiv\sum_{i=1}^n\left( x_i^\alpha \frac{\partial}{\partial x_i^\alpha}\right). 
\end{equation}
 
 Such CWIs are sufficient to completely determine the expression of a scalar 3-point function of primary operators $\mathcal{O}_i$ of scaling dimensions 
$\Delta_i$ $(i=1,2,3)$ in the form
\begin{equation}
\label{corr}
\langle \mathcal{O}_1(\mathbf{x}_1)\mathcal{O}_2(\mathbf{x}_2)\mathcal{O}_3(\mathbf{x}_3)\rangle =\frac{C_{123}}{ x_{12}^{\Delta_t - 2 \Delta_3}  x_{23}^{\Delta_t - 2 \Delta_1}x_{13}^{\Delta_t - 2 \Delta_2} },\qquad \Delta_t\equiv \sum_{i=1}^3 \Delta_i,
\end{equation}
where $x_{ij}=|\mathbf{x}_i - \mathbf{x}_j|$ and  $C_{123}$ is a constant which specifies the CFT. For 4-point functions the same constraints are weaker, and the structure of a scalar correlator is identified modulo an arbitrary function of the two cross ratios 
\begin{equation}
\label{uv}
u(x_i)=\frac{x_{12}^2 x_{34}^2}{x_{13}^2 x_{24}^2} \qquad v(x_i)=\frac{x_{23}^2 x_{41}^2}{x_{13}^2 x_{24}^2}.
\end{equation}
The general solution, allowed by the symmetry, can be written in the form 
\begin{equation}
\label{general}
\langle \mathcal{O}_1(\mathbf{x}_1)\mathcal{O}_2(\mathbf{x}_2)\mathcal{O}_3(\mathbf{x}_3)\mathcal{O}_4(\mathbf{x}_4)\rangle= h(u(x_i),v(x_i))\, \frac{1}{\left(x_{12}^2\right)^\frac{\Delta_1 + \Delta_2}{2}\left(x_{3 4}^2\right)^\frac{\Delta_3 + \Delta_4}{2}},
\end{equation}
where $h(u(x_i),v(x_i))$ remains unspecified. \\
For the analysis of $n$-point function it is sometimes convenient to introduce more general notations. For instance, one may define
\begin{equation}
\label{conv}
\begin{aligned}
 &\Phi(\underline{\mathbf{x}})\equiv \langle \mathcal{O}_1(\mathbf{x}_1)\mathcal{O}_2(\mathbf{x}_2)\ldots \mathcal{O}_n(\mathbf{x}_n)\rangle, && e^{i \underline{\mathbf{p x}}}\equiv e^{i(\mathbf{p}_1\mathbf{x}_1 + \mathbf{p}_2 \mathbf{x}_2 + \ldots \mathbf{p}_n \mathbf{x}_n)}, \\
 &\ul{d p}\equiv dp_1 dp_2 \ldots d p_n, && 
\Phi(\ul{\mathbf{p}})\equiv \langle O_1(\mathbf{p}_1)O_2(\mathbf{p}_2)\ldots O_n(\mathbf{p}_n)\rangle ,\qquad 
\end{aligned}
\end{equation}
where each of the integrations $dp_i\equiv d^d p_i$ are performed on the $d$-dimensional components of the momenta  $\mathbf{p}_i=(p_i^1,p_i^2\ldots p_i^d)$. 
It will also be useful to introduce the total momentum $\mathbf{P}=\sum_{j=1}^{n} \mathbf{p}_j$ characterising a given correlator, which vanishes because of the translational symmetry of the correlator in $\mathbb{R}^d$. \\
The momentum constraint in momentum space is enforced via a delta function $\delta(P)$ in the integrand. For instance, translational invariance of $\Phi(\ul{\mathbf{x}})$ gives 
\begin{equation}
\label{ft1}
\Phi(\ul{\mathbf{x}})=\int \ul{dp}\ \delta(\mathbf{P}) \ e^{i\ul{p x}} \ \Phi(\ul{\mathbf{p}}).
\end{equation}
In general, we recall that for an $n$-point function $\Phi(\underline{\mathbf{x}})$, the condition of translational invariance  
\begin{equation}
\langle 
 \mathcal{O}_1(\mathbf{x}_1)\mathcal{O}_2(\mathbf{x}_2),\ldots, \mathcal{O}_n(\mathbf{x}_n)\rangle = \langle\mathcal{O}_1(\mathbf{x}_1+\mathbf{a} )\mathcal{O}_2(\mathbf{x}_2+\mathbf{a})\ldots \mathcal{O}_n(\mathbf{x}_n+\mathbf{a})\rangle  \end{equation}
 generates an expression of the same correlator in momentum space of the form (\ref{ft1}). We can remove one of the momenta, and conventionally we do it by selecting the last one, $\mathbf{p}_n$, which is replaced by its "on shell" version 
 $\bar{\mathbf{p}}_n=-(\mathbf{p}_1+\mathbf{p}_2 +\ldots +\mathbf{p}_{n-1})$
 \begin{equation}
 \Phi(\ul{\mathbf{x}})=\int dp_1 dp_2... dp_{n-1}e^{i(\mathbf{p}_1 \mathbf{x}_1 + \mathbf{p}_2 \mathbf{x}_2 +...\mathbf{p}_{n-1} \mathbf{x}_{n-1} + 
 \bar{\mathbf{p}}_n \mathbf{x}_n)} \Phi(\mathbf{p}_1,\ldots \mathbf{p}_{n-1},\bar{\mathbf{p}}_n),
 \end{equation}
 denoting with 
 \begin{equation}
 \Phi(\mathbf{p}_1,\ldots \mathbf{p}_{n-1},\bar{\mathbf{p}}_n)=\langle O_1(\mathbf{p}_1)\ldots O_n(\bar{\mathbf{p}}_n)\rangle,
 \end{equation}
 the Fourier transform of the original correlator \eqref{defop}. A discussion of the derivations of the expressions in momentum space of the dilatation and special conformal transformations can be found in \cite{Coriano:2018bbe}. \\
 The special conformal generator in momentum space takes the form
\begin{equation}
K^\kappa(p_i)\equiv\sum_{j=1}^{n-1}\left(2(\Delta_j- d)\frac{\partial}{\partial p_j^\kappa}+p_j^\kappa \frac{\partial^2}{\partial p_j^\alpha\partial p_j^\alpha} -2 p_j^\alpha\frac{\partial^2}{\partial p_j^\kappa \partial p_j^\alpha}\right).
\end{equation}
The latter corresponds to \eqref{transf1}, and then the special CWIs are given  by the equation
\begin{equation}
K^\kappa(p_i)\Phi(\mathbf{p}_1,\ldots \mathbf{p}_{n-1},\bar{\mathbf{p}}_n)=0.\label{SCWI}
\end{equation}

If the primary operator $\mathcal{O}_i$ transforms under a scaling in the form
\begin{equation}
\mathcal{O}_i(\lambda\ \mathbf{x}_i)=\lambda^{-\Delta_i}\mathcal{O}_i(\mathbf{x}_i), 
\end{equation}
in momentum space the same scaling takes the form
\begin{equation}
\Phi(\lambda\,\mathbf{p}_1,\ldots, \lambda\,\bar{\mathbf{p}}_n)=\lambda^{-\Delta'}\Phi(\mathbf{p}_1,\ldots ,\bar{\mathbf{p}}_n),
\end{equation}
with 
\begin{equation}
\Delta'\equiv \left(-\sum_{i=1}^n \Delta_i +(n-1) d\right)=-\Delta_t +(n-1) d.
\end{equation}

In momentum space, the condition of scale covariance and invariance are respectively given by 
\begin{equation}
D(p_i)\,\Phi(\mathbf{p}_1,\ldots ,\bar{\mathbf{p}}_n)=0,
\end{equation}
with
\begin{equation}
D(p_i)\equiv\sum_{i=1}^{n-1}  p_i^\alpha \frac{\partial}{\partial p_i^\alpha} + \Delta'
\end{equation}
and 
\begin{equation}
D_0(p_i)\,\Phi(\mathbf{p}_1,\ldots ,\bar{\mathbf{p}}_n)=0,
\end{equation}
with
\begin{equation}
D_0(p_i)\equiv\sum_{i=1}^{n-1} p_i^\alpha \frac{\partial}{\partial p_i^\alpha} .
\end{equation}
In the case of tensor correlators the structure of the special CWI's involve also the Lorentz generators $\Sigma^{\mu\nu}$ and take the form
\begin{equation}
\label{GenFormSCWI}
\begin{split}
& \sum_{r=1}^{n-1} \left( p_{r \, \mu} \, \frac{\partial^2}{\partial p_{r}^{\nu} \partial p_{r \, \nu}}  - 2 \, p_{r \, \nu} \, 
\frac{\partial^2}{ \partial p_{r}^{\mu} \partial p_{r \, \nu} }    + 2 (\Delta_r - d) \frac{\partial}{\partial p_{r}^{\mu}}  + 2 
(\Sigma_{\mu\nu}^{(r)})^{i_r}_{j_r} \frac{\partial}{\partial p_{r \, \nu}} \right) \\ 
& \hspace{7cm} \, \times \langle \mathcal O^{i_1}_1(\mathbf{p}_1) \ldots  \mathcal O^{j_r}_r(\mathbf{p}_r) \ldots \mathcal O^{i_n}_n(\mathbf{p}_n) \rangle = 0 \, ,
\end{split}
\end{equation}
where the indices $i_1,\ldots i_n$ and $j_1\ldots j_n$ run on the representation of the Lorentz group to which the operators belong. Notice that the sum over the index $r$ selects in each term a specific momentum $p_r$, but the last momentum $p_n$ is not included, since the summation runs from 1 to $n-1$. Therefore the differentiation respect to the last momentum $p_n$, which has been chosen as the dependent one, is performed implicitly. At the same time, the action of the rotation (Lorentz) generators $\Sigma_{\mu\nu}^{(r)}$ of $SO(d)$ is performed on each of the primary operators 
$O_1, O_2\ldots$, except the last one, $O_n$, which is treated like a singlet under such rotational symmetry \cite{Coriano:2018bbe}.

\subsection{2-point functions}
The simplest application of such equations are for 2-point functions \cite{Coriano:2013jba}
$G^{ij}(\mathbf{p}) \equiv \langle \mathcal O_1^i(\mathbf{p}) \mathcal O_2^j(-\mathbf{p}) \rangle$ of two primary fields, each of spin-1, here defined as $i$ and $j$. In this case, if we consider the correlator of two primary fields each of spin-1, the equations take the form
\begin{equation}
\label{ConformalEqMomTwoPoint}
\begin{split}
& \left( - p_{\mu} \, \frac{\partial}{\partial p_{\mu}}  + \Delta_1 + \Delta_2 - d \right) G^{ij}(\mathbf{p}) = 0 \,, \\
& \left(  p_{\mu} \, \frac{\partial^2}{\partial p^{\nu} \partial p_{\nu}}  - 2 \, p_{\nu} \, \frac{\partial^2}{ \partial p^{\mu} 
	\partial p_{\nu} }    + 2 (\Delta_1 - d) \frac{\partial}{\partial p^{\mu}} \right) G^{ij}(\mathbf{p}) + 2 (\Sigma_{\mu\nu})^{i}_{k} \frac{\partial}{\partial 
	p_{\nu}}   G^{kj}(\mathbf{p})  = 0 \,,
\end{split}
\end{equation}
For the 2-point function $G_S(\mathbf{p})$ of two scalar quasi primary fields, the invariance under the Poincar\'e group implies that the function $G_S$ depends on the scalar invariant $p^2$ and then $G_S(\mathbf{p})=G_S(p^2)$. Furthermore, the invariance
under scale transformations implies that $G_S(p^2)$ is a homogeneous function of degree 
$\alpha = \frac{1}{2}(\Delta_1 + \Delta_2 - d)$. It is easy to show that one of the two equations in (\ref{ConformalEqMomTwoPoint}) can be satisfied only if $\Delta_1 = \Delta_2$. 
Therefore conformal symmetry fixes the structure of the scalar two-point function up to an arbitrary overall constant $C$ as
\begin{equation}
\label{TwoPointScalar}
G_S(p^2) = \langle \mathcal O_1(\mathbf{p}) \mathcal O_2(-\mathbf{p}) \rangle = \delta_{\Delta_1 \Delta_2}  \, C\, (p^2)^{\Delta_1 - d/2} \, .
\end{equation}
If we redefine
\begin{equation}
C=c_{S 12} \,  \frac{\pi^{d/2}}{4^{\Delta_1 - d/2}} \frac{\Gamma(d/2 - \Delta_1)}{\Gamma(\Delta_1)} 
\end{equation}
in terms of the new integration constant $c_{S 12}$, the two-point function reads as
\begin{equation}
\label{TwoPointScalar2}
G_S(p^2) =  \delta_{\Delta_1 \Delta_2}  \, c_{S 12} \,  \frac{\pi^{d/2}}{4^{\Delta_1 - d/2}} \frac{\Gamma(d/2 - \Delta_1)}{\Gamma(\Delta_1)} 
(p^2)^{\Delta_1 - d/2} \,,
\end{equation}
and after a Fourier transform in coordinate space takes the familiar form
\begin{equation}
\langle \mathcal O_1(\mathbf{x}_1) \mathcal O_2(\mathbf{x}_2) \rangle \equiv \mathcal{F.T.}\left[ G_S(p^2) \right] =  \delta_{\Delta_1 \Delta_2} \,  c_{S 12} 
\frac{1}{x_{12}^{2\Delta_1}} \,,
\end{equation}
where $x_{12} = |\mathbf{x}_1 - \mathbf{x}_2|$. 
\section{The hypergeometric structure from 3-point functions \texorpdfstring{$F_4$}{F4}} 
In the case of a scalar correlator of 3-point functions, all the conformal WI's can 
be re-expressed in scalar form by taking as independent momenta the magnitude $ {p}_i=|\mathbf{p}_i|=\sqrt{\mathbf{p}_i^2}$. In fact, Lorentz invariance on the correlation function implies that 
\begin{equation}
\Phi(\mathbf{p}_1,\mathbf{p}_2,\bar{\mathbf{p}}_3)=\Phi(p_1,p_2,p_3),\notag
\end{equation}
i.e., it is a function which depends on the magnitude of the momenta $p_i$, $i=1,2,3$. 
In this case $\mathbf{p}_3$ is taken as the dependent momentum ($\bar{\mathbf{p}}_3=-\mathbf{p}_1-\mathbf{p}_2$) by momentum conservation, with $p_3=|\mathbf{p}_1+\mathbf{p}_2|$. The original equations, in the covariant version, take the form 
\begin{equation}
K^\kappa(p_i)\Phi(\mathbf{p}_1,\mathbf{p}_2,\bar{\mathbf{p}}_3)\equiv\sum_{j=1}^{2}\left(2(\Delta_j- d)\frac{\partial}{\partial p_j^\kappa}+p_j^\kappa \frac{\partial^2}{\partial p_j^\alpha\partial p_j^\alpha} -2 p_j^\alpha\frac{\partial^2}{\partial p_j^\kappa \partial p_j^\alpha}\right)\Phi(\mathbf{p}_1,\mathbf{p}_2,\bar{\mathbf{p}}_3)=0,\label{Scw}
\end{equation}
for the special conformal WI and

\begin{equation}
D(p_i)\Phi(\mathbf{p}_1,\mathbf{p}_2,\bar{\mathbf{p}}_3)\equiv\left(\sum_{i=1}^{2}  p_i^\alpha \frac{\partial}{\partial p_i^\alpha} + \Delta'\right)
\Phi(\mathbf{p}_1,\mathbf{p}_2,\bar{\mathbf{p}}_3)
\end{equation}
for the dilatation WI. In this case $K^\kappa(p_i)$ doesn't involve the spin part $\Sigma$, as illustrated in the general expression \eqref{GenFormSCWI}, because of the scalar nature of this particular correlation function. For this reason, the action of $K^\kappa$ is purely scalar  
$K^\kappa(p_i)\equiv K_{scalar}^\kappa(p_i)$.
Using the chain rule
\begin{equation}
\label{chainr}
\frac{\partial \Phi}{\partial p_i^\mu}=\frac{p_i^\mu}{  p_i}\frac{\partial\Phi}{\partial  p_i} 
-\frac{\bar{p}_3^\mu}{  p_3}\frac{\partial\Phi}{\partial   p_3}\qquad i=1,2,
\end{equation}
and the properties of the scalar products
\begin{align}
\mathbf{p}_1\cdot \mathbf{p}_2&=\frac{1}{2}\left[p_3^2-p_1^2-p_2^2\right]\notag\\
\mathbf{p}_i\cdot \mathbf{p}_3&=\frac{1}{2}\left[p_j^2-p_3^2-p_i^2\right], \quad i\ne j,\  i,j=1,2\ ,\notag
\end{align}
one can re-express the differential operator for the dilatation WI as
\begin{equation}
p_1^\alpha \frac{\partial \Phi}{{p_1}^\alpha}+ p_2^\alpha \frac{\partial \Phi}{{p_2}^\alpha}=
{p}_1\frac{ \partial \Phi}{\partial   p_1} +   p_2\frac{ \partial \Phi}{\partial   p_2} +   p_3\frac{ \partial \Phi}{\partial   p_3},
\end{equation}
giving the equation 
\begin{equation}
\label{scale1}
\left(\sum_{i=1}^3\Delta_i -2 d - \sum_{i=1}^3    p_i \frac{ \partial}{\partial   p_i}\right)\Phi(p_1,p_2,p_3)=0.
\end{equation}
One can show that the special conformal transformations, summarised in \eqref{Scw},  
take the form
\begin{equation}
\sum_{i=1}^3 p_i^\kappa \left(\,{K}_i\, \Phi(p_1,p_2,p_3)\right)=0,
\label{kappa2}
\end{equation}
having introduced the operators
\begin{equation}
{ K}_i\equiv \frac{\partial^2}{\partial    p_i \partial    p_i} 
+\frac{d + 1 - 2 \Delta_i}{   p_i}\frac{\partial}{\partial   p_i}.
\end{equation}
It is easy to show that Eq. (\ref{kappa2}) can be split into the two independent equations
\begin{equation}
\frac{\partial^2\Phi}{\partial   p_i\partial   p_i}+
\frac{1}{  p_i}\frac{\partial\Phi}{\partial  p_i}(d+1-2 \Delta_1)-
\frac{\partial^2\Phi}{\partial   p_3\partial   p_3} -
\frac{1}{  p_3}\frac{\partial\Phi}{\partial  p_3}(d +1 -2 \Delta_3)=0\qquad i=1,2,
\label{3k1}
\end{equation}
having used the momentum conservation equation $p_3^\kappa=-p_1^\kappa-p_2^\kappa$. \\
By defining 
\begin{equation}
\label{kij}
K_{ij}\equiv {K}_i-{K}_j,
\end{equation}
Eqs. (\ref{3k1}) take the form 
\begin{equation}
\label{3k2}
K_{13}\,\Phi(p_1,p_2,p_3)=0 \qquad \textrm{and} \qquad K_{23}\,\Phi(p_1,p_2,p_3)=0,
\end{equation}
which are equivalent to a hypergeometric system of equations, with solutions given by linear combinations of Appell's functions $F_4$.

\section{Hypergeometric systems} 
Appell's hypergeometric functions $F_1(x,y)$, $F_2(x,y)$, $F_3(x,y)$, $F_4(x,y)$ are defined by the  hypergeometric series: 
\begin{eqnarray} \label{appf1}
\app 1{a;\;b_1,b_2}{c}{x,\,y} \equal \sum_{n=0}^{\infty} \sum_{m=0}^{\infty}
\frac{(a)_{n+m}\,(b_1)_n\,(b_2)_m}{(c)_{n+m}\;n!\,m!}\,x^n\,y^m,\\ \label{appf2}
\app 2{a;\;b_1,b_2}{c_1,c_2}{x,\,y} \equal \sum_{n=0}^{\infty} \sum_{m=0}^{\infty}
\frac{(a)_{n+m}\,(b_1)_n\,(b_2)_m}{(c_1)_n\,(c_2)_m\;n!\,m!}\,x^n\,y^m,\\ \label{appf3}
\app 3{\!a_1,a_2;\,b_1,b_2}{c}{x,\,y} \equal \sum_{n=0}^{\infty} \sum_{m=0}^{\infty}
\frac{(a_1)_n(a_2)_m(b_1)_n(b_2)_m}{(c)_{n+m}\;n!\,m!}\,x^n\,y^m,\\ \label{appf4}
F_4(a,b,c_1,c_2; x,y)\equiv\app 4{a;\;b}{c_1,c_2\,}{x,\,y} \equal \sum_{n=0}^{\infty} \sum_{m=0}^{\infty}
\frac{(a)_{n+m}\,(b)_{n+m}}{(c_1)_n\,(c_2)_m\;n!\,m!}\,x^n\,y^m
\end{eqnarray}
and are bivariate generalizations of the Gauss hypergeometric series
\begin{equation} \label{gausshpg}
\hpg21{A,\,B}{C}{z} = \sum_{n=0}^{\infty} 
\frac{(A)_{n}\,(B)_n}{(C)_n\,n!}\,z^n.
\end{equation}
with the (Pochhammer) symbol $(\alpha)_{k}$ given by
\begin{equation}
(\alpha)_{k}\equiv (\alpha,k)\equiv\frac{\Gamma(\alpha+k)}{\Gamma(\alpha)}=\alpha(\alpha+1)\dots(\alpha+k-1).\label{Pochh}
\end{equation}
An account of many of the properties of such functions and a discussion of the univariate cases, when the two variables coalesce, can be found in \cite{Vidunas1} and related works. They are solutions of equations generalizing Euler's hypergeometric equation 
\begin{equation} \label{eq:euler}
z(1-z)\,\frac{d^2y(z)}{dz^2}+
\big(C-(A+B+1)z\big)\frac{dy(z)}{dz}-A\,B\,y(z)=0,
\end{equation}
whose solution is denoted as $\hpgo21$, written in (\ref{gausshpg}).
This is classified as a Fuchsian equation with singularities at $z=0$, $z=1$ and $z=\infty$. 
When the two arguments $x,y$ of the Appell functions are algebraically related, they 
are referred to as univariate functions and satisfy Fuchsian ordinary
differential equations.\\
The proof that the CWIs of 3-point functions are hypergeometric systems of equations has been shown independently in \cite{Coriano:2013jba} and \cite{Bzowski:2013sza}. We recall that, in the case of Appell functions of type $F_4$ given in \eqref{appf4}, which are the relevant ones in all our discussion, such functions are solutions of the system of differential equations
\begin{equation}
\label{F4diff.eq}
\begin{cases}
\bigg[ x(1-x) \frac{\partial^2}{\partial x^2} - y^2 \frac{\partial^2}{\partial y^2} - 2 \, x \, y \frac{\partial^2}{\partial x \partial y} +  \left[ \gamma - (\alpha + \beta + 1) x \right] \frac{\partial}{\partial x} \nn \\
\hspace{8cm} - (\alpha + \beta + 1) y \frac{\partial}{\partial y}  - \alpha \, \beta \bigg] F(x,y) = 0 \,, \nn \\
\bigg[ y(1-y) \frac{\partial^2}{\partial y^2} - x^2 \frac{\partial^2}{\partial x^2} - 2 \, x \, y \frac{\partial^2}{\partial x \partial y} +  \left[ \gamma' - (\alpha + \beta + 1) y \right] \frac{\partial}{\partial y} \nn \\
\hspace{8cm} - (\alpha + \beta + 1) x \frac{\partial}{\partial x}  - \alpha \, \beta \bigg] F(x,y) = 0 \,, 
\end{cases} 
\\
\end{equation}
as illustrated in \cite{Appell}, where $F(x,y)$ can be in the most general case a linear combinations of 4 independent functions $F_4$, hypergeometric of two variables $x$ and $y$. The univariate limits of the solutions are important, from the physical point of view, for the study of the behaviour of the corresponding correlation functions in special kinematics. An example has been discussed in \cite{Coriano:2019nkw} in the case of 4-point functions. 

\subsection{Scalar 3-point functions}
To show the emergence of such system of equations, let's focus on a 3-point function of three primary scalar fields of a generic CFT  
\begin{equation}
\langle \mathcal O_1(\mathbf{p}_1) \mathcal O_2(\mathbf{p}_2) \mathcal O_3(\bar{\mathbf{p}}_3) \rangle \,=\Phi(p_1,p_2,p_3),
\end{equation}
 defined by the two homogeneous conformal equations
\begin{equation}
K_{31}\Phi=0  \qquad K_{21}\Phi=0,
\end{equation}
combined with the scaling equation 
\begin{equation}
\label{scale}
\sum_{i=1}^3 p_i\frac{\partial}{\partial p_i} \Phi=(\Delta-2 d) \Phi, 
\end{equation}
where $\Delta=\Delta_1+\Delta_2+\Delta_3$.
We follow the analysis of in \cite{Coriano:2013jba} and introduce the ansatz
\begin{equation}
\label{ans}
\Phi(p_1,p_2,p_3)=p_1^{\Delta - 2 d} x^{a}y^{b} F(x,y),
\end{equation}

with $x=\frac{p_2^2}{p_1^2}$ and $y=\frac{p_3^2}{p_1^2}$. One of the three momenta is treated asymmetrically and takes the role of "the pivot" in the representation of the function as a series. In this case we have chosen $p_1$ as a pivot, but we could have equivalently chosen any of the 3  momenta. \\
$\Phi$ is required to be homogenous of degree $\Delta-2 d$ under a scale transformation, according to (\ref{scale}), and in (\ref{ans}) this is taken into account by the factor $p_1^{\Delta - 2 d}$.
The use of the scale invariant variables $x$ and $y$ takes to the hypergeometric form of the solution. One obtains by an implicit differentiation the identity
\begin{align}
K_{21}\phi &= 4 p_1^{\Delta -2d -2} x^a y^b
\left(  x(1-x)\frac{\partial }{\partial x \partial x}  + (A x + \gamma)\frac{\partial }{\partial x} -
2 x y \frac{\partial^2 }{\partial x \partial y}- y^2\frac{\partial^2 }{\partial y \partial y} + 
D y\frac{\partial }{\partial y} + \left(E +\frac{G}{x}\right)\right) \notag\\
& \hspace{3cm}\times F(x,y)=0,
\label{red}
\end{align}
with
\begin{align}
&A=D=\Delta_2 +\Delta_3 - 1 -2 a -2 b -\frac{3 d}{2} \qquad \gamma(a)=2 a +\frac{d}{2} -\Delta_2 + 1
\notag\\
& G=\frac{a}{2}(d +2 a - 2 \Delta_2)
\notag\\
&E=-\frac{1}{4}(2 a + 2 b +2 d -\Delta_1 -\Delta_2 -\Delta_3)(2 a +2 b + d -\Delta_3 -\Delta_2 +\Delta_1).
\end{align}
The treatment of equation $K_{31}\Phi=0$ proceeds in a similar way, with the obvious exchanges $(a,b,x,y)\to (b,a,y,x)$
\begin{align}
K_{31}\phi &= 4 p_1^{\Delta -2 d -2} x^a y^b
\left(  y(1-y)\frac{\partial }{\partial y \partial y}  + (A' y + \gamma')\frac{\partial }{\partial y} -
2 x y \frac{\partial^2 }{\partial x \partial y}- x^2\frac{\partial^2 }{\partial x \partial x} + 
D' x\frac{\partial }{\partial x} + \left(E' +\frac{G'}{y}\right)\right) \notag\\
& \hspace{3cm}\times F(x,y)=0,
\end{align}
with
\begin{align}
&A'=D'= A   \qquad \qquad \gamma'(b)=2 b +\frac{d}{2} -\Delta_3 + 1
\notag\\
& G'=\frac{b}{2}(d +2 b - 2 \Delta_3)
\notag\\
&E'= E.
\end{align}
Eq. (\ref{red}) acquires a hypergeometric form if we set $G/x=0$, which implies that
\begin{equation}
\label{cond1}
a=0\equiv a_0 \qquad \textrm{or} \qquad a=\Delta_2 -\frac{d}{2}\equiv a_1.
\end{equation}
The equation $K_{31}\Phi=0$ generates a similar condition for $b$ by setting $G'/y=0$, fixing the two remaining indices
\begin{equation}
\label{cond2}
b=0\equiv b_0 \qquad \textrm{or} \qquad b=\Delta_3 -\frac{d}{2}\equiv b_1.
\end{equation}
Notice that the elimination of such a singularity in the equations guarantees, from the physical perspective, the analyticity of the solutions and the absence of unphysical thresholds which are not expected in a massless theory.
The four independent solutions of the CWI's will all be characterised by the same 4 pairs of indices $(a_i,b_j)$ $(i,j=1,2)$. It is convenient to define the two functions 
\begin{equation}
\alpha(a,b)= a + b + \frac{d}{2} -\frac{1}{2}(\Delta_2 +\Delta_3 -\Delta_1) \qquad \beta (a,b)=a +  b + d -\frac{1}{2}(\Delta_1 +\Delta_2 +\Delta_3), \qquad 
\label{alphas}
\end{equation}
then
\begin{equation}
E=E'=-\alpha(a,b)\beta(a,b) \qquad A=D=A'=D'=-\left(\alpha(a,b) +\beta(a,b) +1\right).
\end{equation}
Therefore, the solution takes  the hypergeometric form in which the functions above take the role of parameters
\begin{align}
\label{F4def}
F_4(\alpha(a,b), \beta(a,b); \gamma(a), \gamma'(b); x, y) = \sum_{i = 0}^{\infty}\sum_{j = 0}^{\infty} \frac{(\alpha(a,b), {i+j}) \, 
	(\beta(a,b),{i+j})}{(\gamma(a),i) \, (\gamma'(b),j)} \frac{x^i}{i!} \frac{y^j}{j!}. 
\end{align}
 We will refer to $\alpha\ldots \gamma'$ as to the first,$\ldots$, fourth parameters of $F_4$.\\ 
Notice that the system is of order 4, since it allows 4 independent solutions which are then all of the form $x^a y^b F_4$, where the 
hypergeometric functions will take some specific values for its parameters, with
$a$ and $b$ fixed by (\ref{cond1}) and (\ref{cond2})
\begin{equation}
\Phi(p_1,p_2,p_3)=p_1^{\Delta-2 d} \sum_{a,b} c(a,b,\vec{\Delta})\,x^a y^b \,F_4(\alpha(a,b), \beta(a,b); \gamma(a), \gamma'(b); x, y). 
\label{compact}
\end{equation}
In the expression above the sum runs over the four values $a_i, b_i$ $i=0,1$ with arbitrary constants $c(a,b,\vec{\Delta})$, with $\vec{\Delta}=(\Delta_1,\Delta_2,\Delta_3)$. The sum over $a$ and $b$ 
needs to be made explicit and, at the same time, one has to combine the 4 independent solutions in such a way that the symmetry respect to the three external momenta is restored. 
For this reason if we define
\begin{align}
&\alpha_0\equiv \alpha(a_0,b_0)=\frac{d}{2}-\frac{\Delta_2 + \Delta_3 -\Delta_1}{2},\, && \beta_0\equiv \beta(b_0)=d-\frac{\Delta_1 + \Delta_2 +\Delta_3}{2},  \nn
&\gamma_0 \equiv \gamma(a_0) =\frac{d}{2} +1 -\Delta_2,\, &&\gamma'_0\equiv \gamma(b_0) =\frac{d}{2} +1 -\Delta_3,
\end{align}
to be the 4 basic hypergeometric parameters, with the remaining ones determined by shifts respect to these values, the four fundamental solutions can be expressed in the form

\begin{align}
\label{sol}
S_1(\alpha_0, \beta_0; \gamma_0, \gamma'_0; x, y)\equiv F_4(\alpha_0, \beta_0; \gamma_0, \gamma'_0; x, y) = \sum_{i = 0}^{\infty}\sum_{j = 0}^{\infty} \frac{(\alpha_0,i+j) \, 
	(\beta_0,i+j)}{(\gamma_0,i )\, (\gamma'_0,j)} \frac{x^i}{i!} \frac{y^j}{j!},
\end{align}
valid for $\sqrt{ x} +\sqrt{y} <1$ together with
\bea
\label{solutions}
S_2(\alpha_0, \beta_0; \gamma_0, \gamma'_0; x, y) &=& x^{1-\gamma_0} \, F_4(\alpha_0-\gamma_0+1, \beta_0-\gamma_0+1; 2-\gamma_0, \gamma'_0; x,y) \,, \nn \\
S_3(\alpha_0, \beta_0; \gamma_0, \gamma'_0; x, y) &=& y^{1-\gamma'_0} \, F_4(\alpha_0-\gamma'_0+1,\beta_0-\gamma'_0+1;\gamma_0,2-\gamma'_0 ; x,y) \,, \nn \\
S_4(\alpha_0, \beta_0; \gamma_0, \gamma'_0; x, y) &=& x^{1-\gamma_0} \, y^{1-\gamma'_0} \, 
F_4(\alpha_0-\gamma_0-\gamma'_0+2,\beta_0-\gamma_0-\gamma'_0+2;2-\gamma_0,2-\gamma'_0 ; x,y) \, . \nn
\eea
The symmetrization with respect to the external momenta \cite{Coriano:2013jba}, by 
using the formula 
\begin{align}
\label{transfF4}
F_4(\alpha, \beta; \gamma, \gamma'; x, y) =& \quad\frac{\Gamma(\gamma') \Gamma(\beta - \alpha)}{ \Gamma(\gamma' - \alpha) \Gamma(\beta)} (- y)^{- \alpha} \, F_4\left(\alpha, \alpha -\gamma' +1; \gamma, \alpha-\beta +1; \frac{x}{y}, \frac{1}{y}\right) \notag\\ 
&+  \frac{\Gamma(\gamma') \Gamma(\alpha - \beta)}{ \Gamma(\gamma' - \beta) \Gamma(\alpha)} (- y)^{- \beta} \, F_4\left(\beta -\gamma' +1, \beta ; \gamma, \beta-\alpha +1; \frac{x}{y}, \frac{1}{y}\right) \,
\end{align}
allows to reverse the ratios of the momenta in the expansion, and reduces the four constants to just one. 
The solution can then be written in the final form
\begin{equation}
\Phi(p_1,p_2,p_3)=p_3^{\Delta-2 d} \sum_{i=1}^4 \,c_i(\D_1,\D_2,\D_3)\,S_i (\alpha, \beta; \gamma, \gamma'; x, y),
\end{equation}
where $c_i$ are arbitrary coefficients which may depend on the scale dimensions $\D_i$ and on the spacetime dimension $d$. \\
By using \eqref{transfF4}, the general symmetric solution can be expressed in the form 
{\cite{Coriano:2013jba}
	\begin{align}
	&\braket{O(p_1)\,O(p_2)\,O(p_3)}=\big(p_3^2\big)^{-d+\frac{\Delta_t}{2}}\,C(\Delta_1,\Delta_2,\Delta_3,d)\notag\\
	&\Bigg\{\Gamma\left(\Delta_1-\frac{d}{2}\right)\Gamma\left(\Delta_2-\frac{d}{2}\right)\Gamma\left(d-\frac{\Delta_1+\Delta_2+\Delta_3}{2}\right)\Gamma\left(d-\frac{\Delta_1+\Delta_2-\Delta_3}{2}\right)\notag\\
	&\hspace{3cm}\times
	\,F_4\,\left(\frac{d}{2}-\frac{\Delta_1+\Delta_2-\Delta_3}{2},d-\frac{\Delta_t}{2},\frac{d}{2}-\Delta_1+1,\frac{d}{2}-\Delta_2+1;x,y\right)\notag
		\end{align}
	\begin{align}
	&\qquad+\,
	\Gamma\left(\frac{d}{2}-\Delta_1\right)\Gamma\left(\Delta_2-\frac{d}{2}\right)\Gamma\left(\frac{\Delta_1-\Delta_2+\Delta_3}{2}\right)\Gamma\left(\frac{d}{2}+\frac{\Delta_1-\Delta_2-\Delta_3}{2}\right)\notag\\
	&\hspace{3cm}\times x^{\Delta_1-\frac{d}{2}}\,F_4\,\left(\frac{\Delta_1-\Delta_2+\Delta_3}{2},\frac{d}{2}-\frac{\Delta_2+\Delta_3-\Delta_1}{2},\Delta_1-\frac{d}{2}+1,\frac{d}{2}-\Delta_2+1;x,y\right)\notag\\[2ex]\notag
	\end{align}
	\begin{align}
	&\qquad+\,
	\Gamma\left(\Delta_1-\frac{d}{2}\right)\Gamma\left(\frac{d}{2}-\Delta_2\right)\Gamma\left(\frac{-\Delta_1+\Delta_2+\Delta_3}{2}\right)\Gamma\left(\frac{d}{2}+\frac{-\Delta_1+\Delta_2-\Delta_3}{2}\right)\notag\\
	&\hspace{3cm}\times\,y^{\Delta_2-\frac{d}{2}}\,F_4\,\left(\frac{\Delta_2-\Delta_1+\Delta_3}{2},\frac{d}{2}-\frac{\Delta_1-\Delta_2+\Delta_3}{2},\frac{d}{2}-\Delta_1+1,\Delta_2-\frac{d}{2}+1;x,y\right)\notag\\[2ex]
	&\qquad+\,
	\Gamma\left(\frac{d}{2}-\Delta_1\right)\Gamma\left(\frac{d}{2}-\Delta_2\right)\Gamma\left(\frac{\Delta_1+\Delta_2-\Delta_3}{2}\right)\Gamma\left(-\frac{d}{2}+\frac{\Delta_1+\Delta_2+\Delta_3}{2}\right)\notag\\
	&\hspace{3cm}\times\,x^{\Delta_1-\frac{d}{2}}\,y^{\Delta_2-\frac{d}{2}}F_4\,\left(-\frac{d}{2}+\frac{\Delta_t}{2},\frac{\Delta_1+\Delta_2-\Delta_3}{2},\Delta_1-\frac{d}{2}+1,\Delta_2-\frac{d}{2}+1;x,y\right)\Bigg\}.\label{solfin}
	\end{align}
	It is important to verify that the symmetric solution above does not have any unphysical singularity in the physical region, reproducing the expected behaviour in the large momentum limit $p_3\gg p_1$ \cite{Bzowski:2014qja}. Indeed the previous expression, in the limit $p_3\gg p_1$ (expressible also as $p_3^2,\,p_2^2\to\infty$ with $p_2^2/p_3^2\to1$ fixed), behaves as
	\begin{align}
	\braket{O(p_1)\,O(p_2)\,O(p_3)}\propto f(d,\Delta_i)\,p_3^{\Delta_1+\Delta_2+\Delta_3-2d}\left( 1 +O\left(p_1/p_3\right)\right)  \hspace{2cm}\text{if}\ \ \Delta_1>\frac{d}{2}&
	\end{align}
	and
	\begin{align}
	\braket{O(p_1)\,O(p_2)\,O(p_3)}\propto g(d,\Delta_i)\,p_3^{\Delta_2+\Delta_3-\Delta_1-d}\,p_1^{2\Delta_1-d}\left(1 +O\left(p_1/p_3\right)\right)\hspace{2cm} \text{if}\ \ \Delta_1<\frac{d}{2}&,
	\end{align}
	with $f(d,\Delta_i)$ and $g(d,\Delta_i)$ depending only on the scaling and spacetime dimensions. In the case of a scalar 3-point function, the CFT correlator is equivalent to a Feynman master integral, as one can immediately realize.  
	The result above in \eqref{solfin} is in complete agreement with the direct computation performed by Davydychev \cite{Davydychev:1992xr} of such integrals, obtained by a Fourier transform of \eqref{corr} and the use of the Mellin-Barnes method. \\
	An equivalent version of the solution found above can be derived as in \cite{Bzowski:2013sza}, where it is written in terms of $K$ Bessel functions as
	\begin{equation}
	\label{caz}
	\Phi(p_1,p_2,p_3)=\,C_{123}\, p_1^{\D_1-\frac{d}{2}}p_2^{\D_2-\frac{d}{2}}p_3^{\D_3-\frac{d}{2}}\int_0^\infty dx\,x^{\frac{d}{2}-1}\,K_{\D_1-\frac{d}{2}}(p_1\,x)\,K_{\D_2-\frac{d}{2}}(p_2\,x)\,K_{\D_3-\frac{d}{2}}(p_3\,x),
	\end{equation}
	where $C_{123}$ is an undetermined constant. Some details are given in the appendix.
The 3K integral 
\begin{equation}
	 \int_0^\infty d s \: s^{\alpha - 1} K_\lambda(p_1 s) K_\mu(p_2 s) K_\nu(p_3 s) = \frac{2^{\alpha - 4}}{c^\alpha} \left[ B(\lambda, \mu) + B(\lambda, -\mu) + B(-\lambda, \mu) + B(-\lambda, -\mu) \right], \label{3k}
\end{equation}
is related to the hypergeometric functions 	
	\begin{align}
	B(\lambda, \mu) & = \left( \frac{a}{c} \right)^\lambda \left( \frac{b}{c} \right)^\mu \Gamma \left( \frac{\alpha + \lambda + \mu - \nu}{2} \right) \Gamma \left( \frac{\alpha + \lambda + \mu + \nu}{2} \right) \Gamma(-\lambda) \Gamma(-\mu) \times \notag\\
	& \qquad \times F_4 \left( \frac{\alpha + \lambda + \mu - \nu}{2}, \frac{\alpha + \lambda + \mu + \nu}{2}; \lambda + 1, \mu + 1; \frac{a^2}{c^2}, \frac{b^2}{c^2} \right), \label{3Kplus}
	\end{align}
	valid for
	\begin{equation}
	\Re\, \alpha > | \Re\, \lambda | + | \Re \,\mu | + | \Re\,\nu |, \qquad \Re\,(a + b + c) > 0. \nn
	\end{equation}
 We also recall that the Bessel functions $K_\nu$ satisfy the equations 
	\begin{align}
	\frac{\partial}{\partial p}\big[p^\b\,K_\b(p\,x)\big]&=-x\,p^\b\,K_{\b-1}(p x)\nn
	K_{\b+1}(x)&=K_{\b-1}(x)+\frac{2\b}{x}K_{\b}(x), \label{der}
	\end{align}
	which are important in order to verify that \eqref{caz} satisfies the CWI's. 
	It will be convenient to adopt the notation introduced in \cite{Bzowski:2013sza} and define the general expression
	\begin{equation}
	\label{trekappa}
	I_{\a\{\b_1,\b_2,\b_3\}}(p_1; p_2\; p_3)=\int_0^\infty\,dx\,x^\a\,(p_1)^{\b_1}\,(p_2)^{\b_2}\,(p_3)^{\b_3}\,K_{\b_1}(p_1\,x)\,K_{\b_2}(p_2\,x)\,K_{\b_3}(p_3\,x),
	\end{equation}
	which will turn useful in the analysis of scalar 4-point functions.\\
	As we are going to see, a similar form of the solution is obtained in the case of dual-conformal/conformal solutions, where the conformal symmetry in coordinate (or, equivalently, momentum) space, is accompanied by an additional symmetry in a space of dual variables. 
	In perturbative quantum field theory, this symmetry goes under the name of {\em dual conformal}. If we impose on a 4-point function both symmetries, then the CWIs alone are sufficient to fix the solution of the conformal constraints modulo a single overall constant, with no need of introducing any extra formalism, such as the operator product expansion, which is necessary in order to determine the structure of the 4-point functions in a CFT.  
	The method to extract such dual conformal/conformal solutions is quite involved technically and has been developed by us in 
	\cite{Maglio:2019grh}.   \\
	Before coming to a discussion of this point, we illustrate the generalisation of such systems of equations obtained in the scalar case, to tensor correlators, bringing as a nontrivial example the $TJJ$, and showing how the inhomogeneous systems of equations generated by the CWI's can be solved. \\
	We will follow the approach formulated in \cite{Coriano:2018bbe}, which exploits the properties of the single function $F_4$ in order to solve the corresponding systems of equations. The original approach for the solution of these systems of equations, entirely based on the 3K integrals of \eqref{trekappa} has been developed in \cite{Bzowski:2013sza}. The two approaches can be mapped one to the other.\\
	
	\subsection{Tensor correlators: the TJJ} 
	The $TJJ$ is a tensor correlator involving one stress energy tensor $T$ and two vector currents $J$. \\
	The interest in this correlator is manifold since it describes the leading contribution to the coupling to gravity of an ordinary gauge theory, such as quantum electrodynamics (QED) and, in the nonabelian case, quantum chromodynamics (QCD). \\
	Perturbative studies of this correlator have been necessary in order to uncover the manifestation of the conformal (Weyl) anomaly in conformal field theory \cite{Duff:1993wm}, which induces the breaking of a classical symmetry at the classical level. The conformal anomaly is associated with a specific functional which involves both a gravitational part and a gauge part. More details about this topic can be found in a sequel of analysis \cite{Coriano:2017mux, Giannotti:2008cv,Coriano:2013jba}, and in the review \cite{Coriano:2019dyc}. The $TJJ$ is sensitive to the gauge part of the anomaly
	\begin{equation}
	\label{confan}
	\langle T^\mu_\mu \rangle_{A, g}=\beta(\alpha) F_{\alpha \beta} F^{\alpha \beta},
	\end{equation}
	where $\beta(\alpha)$ is the $\beta$ function of the theory which describes the running of the gauge coupling $\alpha$ under a change of the renormalisation scale. In \eqref{confan} we have performed a quantum average $(\langle\,\rangle)$ in the background of two classical fields, the metric $g_{\mu\nu}$ and the gauge field $A_{\mu}$, as defined in terms of a functional integral (see \cite{Coriano:2018bsy}).\\
$F_{\alpha \beta}=\partial_\alpha A_\beta -\partial_\beta A_\alpha$ is the field strength of the abelian gauge field $A_\mu$. Therefore, the classical result of a vanishing trace of $T$ $(T^\mu_\mu=0)$, which is expected due to conformal symmetry, is modified as in \eqref{confan}.  More details on this point can be found in \cite{Coriano:2013jba,Giannotti:2008cv,Armillis:2009pq,Coriano:2018bbe}. \\
The $TJJ$ correlator is defined by the expression 
	\begin{equation}
	\Gamma^{\mu,\nu\alpha\beta}=\langle T^{\mu\nu}(\mathbf{x}_1)J^\alpha(\mathbf{x}_2) J^\beta(\mathbf{x}_3) \rangle
	\end{equation}
	and, as we have mentioned, is affected by an anomaly. This is generated when all the points $x_i$ of the correlator coalesce, and it is for this reason that the analysis of these types of correlators requires a regularisation procedure. \\
	Their analysis can be drastically simplified if, for a given CFT characterised by specific conformal data, we can find a realisation of the same correlator in terms of a specific free field theory. There is one simple strategy in order to perform such mapping. 
	
	1. We need to solve all the CWIs consistently, identifying all the independent constants that appear in the solution. For instance, in the case of the $TJJ$, there will be three undetermined constants. The solution obtained is expressed in terms of linear combinations of functions $F_4$ or of 3K integrals.
	
	2. We consider three independent conformal free field theories, defined by ordinary Lagrangians, with an arbitrary particle content, and compute the Feynman contributions corresponding to the $TJJ$ vertex. These amount to one-loop diagrams and define the perturbative solution. Such solution will be characterised by 
	three independent constants, which can be chosen to be the number of scalars $n_s$, fermions $n_F$ and spin-1 (gauge) fields $n_V$, all running in the loops.
	
	3. The match between the two results can be explicitly checked by going to special spacetime dimensions where the general hypergeometric solutions simplify. This occurs in $d=3$ and $d=5$, which are sufficient to establish a direct match between the three constants of the CWI' and the free field theory realisation $(n_S, n_V, n_F)$ bi-univocally.
	
	As a result of the match, the hypergeometric solutions can be re-expressed in terms of simple scalar massless Feynman integrals, corresponding to the 2- and 3-point functions. 
	Feynman amplitudes corresponding to 3-point functions are indeed described by generalised hypergeometric functions, and this is at the root of the success of the correspondence between the two approaches.\\
	The formulation of the general method which takes to the solution of the CWI's can be found in \cite{Bzowski:2013sza}. It is based on the decomposition of a tensor correlator in terms of its transverse traceless and longitudinal sectors. A similar decomposition is performed on the CWI's. In particular, the hypergeometric structure of the equations emerges from the transverse traceless sector, in a way which is quite close, though more involved, respect to the scalar case discussed above.  \\
	The entire solution is parameterised by a certain number of form factors $A_i$ which are functions of the momentum variables $p_i^2$, appearing in the parameterisation of the transverse traceless sector of the correlator. In the case of the $TJJ$ the equations take the form

	\begin{equation}
	\begin{split}
	K_{13}A_1&=0\\
	K_{13}A_2+2A_1&=0\\
	K_{13}A_3-4A_1&=0\\
	K_{13}A_3(p_2\leftrightarrow p_3)&=0\\
	K_{13}A_4-2A_3(p_2\leftrightarrow p_3)&=0
	\end{split}
	\hspace{1.5cm}
	\begin{split}
	K_{12}A_1&=0\\
	K_{12}A_2+2A_1&=0\\
	K_{12}A_3&=0\\
	K_{12}A_3(p_2\leftrightarrow p_3)-4A_1&=0\\
	K_{12}A_4-2A_3&=0
	\end{split}\label{Primary}
	\end{equation}
	and represent a hypergeometric system of equations which generalizes the simpler result presented in the scalar case.

	\subsection{Solving inhomogeneous systems in the TJJ case}
	We illustrate our method of solving the system of equations above, based on the shifts of the parameters of each function $F_4$. Notice that in this case, the correlator is symmetric respect to the exchange of the two $J$ operators.\\
	As in the case discussed above, we take as a pivot $p_1^2$, and assume the symmetry under the $(P_{23})$ exchange of $(p_2,\Delta_2)$ with $(p_3,\Delta_3)$ in the correlator. In the case of two photons 
	$\Delta_2=\Delta_3=d-1$.
	
	We start from $A_1$ by solving the two equations from (\ref{Primary}) 
	\begin{equation}
	K_{21}A_1=0   \qquad K_{31}A_1=0.
	\end{equation}
	In this case we introduce the ansatz 
	\begin{equation}
	A_1=p_1^{\Delta-2 d - 4}x^a y^b  F(x,y)
	\end{equation}
	where $\Delta=\Delta_1+\Delta_2+\Delta_3$ and derive two hypergeometric equations, which are characterized by the same indices 
	$(a_i, b_j)$ as before in (\ref{cond1}) and (\ref{cond2}), but new values of the 4 defining parameters. 
	We obtain 
	\begin{equation}
	\label{A1}
	A_1(p_1,p_2,p_3)=p_1^{\Delta-2 d - 4}\sum_{a,b} c^{(1)}(a,b,\vec{\Delta})\,x^a y^b \,F_4(\alpha(a,b) +2, \beta(a,b)+2; \gamma(a), \gamma'(b); x, y) 
	\end{equation}
	($\vec{\Delta}\equiv(\Delta_1,\Delta_2,\Delta_3)$), with the expression of $\alpha(a,b),\beta(a,b), \gamma(a), \gamma'(b)$ as given before, with the obvious switching of the 
	$\Delta_i$ in order to comply with the new choice of the pivot ($p_1^2$)
	\begin{align}
	\label{cons1}
	\alpha(a,b)&= a + b + \frac{d}{2} -\frac{1}{2}(\Delta_2 + \Delta_3 -\Delta_1) \notag\\
	\beta(a,b)&= a + b + d -\frac{1}{2}(\Delta_1 + \Delta_2 +\Delta_3) 
	\end{align}
	which are $P_{23}$ symmetric and 
	\begin{align} 
	\label{cons2}
	\gamma(a)& =2 a +\frac{d}{2} -\Delta_2 + 1 \notag\\
	\gamma'(b)&=2 b +\frac{d}{2} -\Delta_3 + 1 
	\end{align}
	with $P_{23}\gamma(a)=\gamma'(b)$.
	If we require that $\Delta_2=\Delta_3$, as in the $TJJ$ case, the symmetry constraints are easily implemented.  The 4 indices, if we choose $p_1$ as a pivot, are given by 
	\begin{equation}
	a_0=0, b_0=0, a_1=\Delta_2- \frac{d}{2}, b_1=\Delta_3-\frac{d}{2} 
	\end{equation}
	and in this case $a=b$ and $\gamma(a)=\gamma(b)$. 
	We recall that $F_4$ has the symmetry 
	\begin{align}
	F_4(\a,\b; \g, \g' ; x,y)=F_4(\a,\b; \g', \g ; y,x),
	\end{align}
	and this reflects in the Bose symmetry of $A_1$ if we impose the constraint
	\be
	c^{(1)}(a_1,b_0,\vec{\Delta})=c^{(1)}(a_0,b_1,\vec{\Delta}).
	\ee
	Now let's turn to the solution for $A_2$.\\
	The equations for $A_2$ are inhomogeneous
	\begin{align}
	K_{21}A_2 &= 2 A_1\label{inhom}\\
	K_{31}A_2& = 2A_1.\label{inhom1}
	\end{align}
	We consider an ansatz of the form 
	\begin{equation}
	A_2(p_1,p_2,p_3)=p_1^{\Delta-2 d - 2}F(x,y)
	\end{equation}
	which provides the correct scaling dimensions for $A_2$. At this point, the action of $K_{21}$ and $K_{31}$ on $A_2$ can be rearranged as follows
	\begin{align}
	K_{21} A_2&=4 x^a y^b p_1^{\Delta-2 d -4}\bigg( \bar{K}_{21}F(x,y) +\frac{\partial}{\partial x} F(x,y)\bigg)\\[1.5ex]
	K_{31} A_2&=4 x^a y^b p_1^{\Delta-2 d -4}\bigg( \bar{K}_{31}F(x,y) +\frac{\partial}{\partial y} F(x,y)\bigg)
	\end{align}
	where
	\begin{align}
	\label{k1bar}
	\bar{K}_{21}F(x,y)&=\bigg\{x(1-x) \frac{\partial^2}{\partial x^2} - y^2 \frac{\partial^2}{\partial y^2} - 2 \, x \, y \frac{\partial^2}{\partial x \partial y} +\big[  (\gamma(a)-1) - (\alpha(a,b) + \beta(a,b) + 3) x \big] \frac{\partial}{\partial x}\notag\\
	&\hspace{3cm}
	+ \frac{a (a-a_1)}{x} - (\alpha(a,b) + \beta(a,b) + 3) y \frac{\partial}{\partial y}  - (\alpha +1)(\beta +1) \bigg\} F(x,y),
	\end{align}
	and 
	\begin{align}
	\label{k2bar}
	\bar{K}_{31} A_2&=\bigg\{ y(1-y) \frac{\partial^2}{\partial y^2} - x^2 \frac{\partial^2}{\partial x^2} - 2 \, x \, y \frac{\partial^2}{\partial x \partial y} +  \big[ (\gamma'(b)-1)- (\alpha(a,b) + \beta(a,b) + 3) y \big]\frac{\partial}{\partial y}\notag\\
	&\hspace{2cm} +  \frac{b(b- b_1)}{y} - (\alpha(a,b) + \beta(a,b) + 3) x \frac{\partial}{\partial x}  -(\alpha(a,b) +1)(\beta(a,b) +1)\bigg\} F(x,y),
	\end{align}
	and we notice that the hypergeometric function which solves the equation
	\begin{equation}
	\label{ffirst1}
	\bar{K}_{21}F(x,y)=0
	\end{equation}
	can be taken of the form
	\begin{equation}
	\label{rep1}
	\Phi_1^{(2)}(x,y)=p_1^{\Delta-2 d - 2}\sum_{a,b} c^{(2)}_1(a,b,\vec{\Delta})\,x^a y^b \,F_4(\alpha(a,b) +1, \beta(a,b)+1; \gamma(a)-1, \gamma'(b) ; x, y) 
	\end{equation}
	with $c^{(2)}_1$ a constant and the parameters $a,b$ fixed at the ordinary values $(a_i,b_j)$ as in the previous cases (\ref{cond1}) and (\ref{cond2}). This allow to remove the $1/x$ and $1/y$ poles in the coefficients of the differential operators. The sequence of parameters in (\ref{rep1}) will obviously solve the related equation 
	\begin{equation}
	\label{sec}
	{K}_{31}\Phi_1^{(2)}(x,y)=0. 
	\end{equation}
	Eq. (\ref{ffirst1}) can be verified by observing that the sequence of  parameters 
	$(\alpha(a,b)+1,\beta(a,b)+1 \gamma(a)-1)$ allows to define a solution of (\ref{k2bar}) set to zero, for an arbitrary $\gamma'(b)$, since this parameter does not play any role in the solution of the corresponding equation. 
	It is also clear that the sequence $(\alpha(a,b)+1,\beta(a,b)+1, \gamma'(b))$, on the other hand, solves the homogeneous equations associated to $K_{31}$ (i.e.,Eq. (\ref{sec})) for any value of the third parameter of $F_4$, which in this case takes the value $\gamma(a)-1$.
	We can follow the same approach for the mirror solution
	\begin{equation} 
	\label{rep}
	\Phi_2^{(2)}(x,y)= p_1^{\Delta-2 d - 2}\sum_{a,b} c_2^{(2)}(a,b,\vec{\Delta})\,x^a y^b \,F_4(\alpha(a,b) +1, \beta(a,b)+1; \gamma(a), \gamma'(b)-1 ; x, y) 
	\end{equation}
	which satisfies 
	\begin{equation}
	\bar{K}_{31}\Phi_2^{(2)}(x,y)=0 \qquad  {K}_{21}\Phi_2^{(2)}(x,y)=0. 
	\end{equation}
	As previously remarked, the values of the exponents $a$ and $b$ remain the same for any equation involving either a $K_{i,j}$ or a $\bar{K}_{i j}$, as can be explicitly verified. This clearly shows that the fundamental solutions of the conformal equations are essentially given by the 4 functions of the type $S_1,\ldots S_4$, for appropriate values of their parameters.  \\
	At this point we use the property 
	\begin{equation}
	\frac{\partial^{p+q} F_4(\alpha,\beta;\gamma_1,\gamma_2;x,y)}{\partial x^p\partial y^q} =\frac{(\alpha,p+q)(\beta,p+q)}{(\gamma_1,p)(\gamma_2,q)}
	F_4(\alpha + p + q,\beta + p + q; \gamma_1 + p ; \gamma_2 + q;x,y)
	\end{equation}
	which gives (for generic parameters $\alpha,\beta,\gamma_1,\gamma_2$)
	\begin{align}
	& \frac{\partial F_4(\alpha,\beta;\gamma_1,\gamma_2;x,y)}{\partial x} =\frac{\alpha \beta}{\gamma_1}F_4(\alpha+1,\beta+1,\gamma_1+1,\gamma_2,x,y) \notag\\
	&  \frac{\partial F_4(\alpha,\beta;\gamma_1,\gamma_2;x,y)}{\partial y} =\frac{\alpha \beta}{\gamma_2}F_4(\alpha+1,\beta+1,\gamma_1,\gamma_2 +1,x,y).
	\end{align} 
	Obviously, such relations are valid independently of the four parameters $\alpha,\beta,\gamma_1,\gamma_2$. 
	The actions of $K_{21}$ and $K_{31}$ on the the $\Phi_2^{(i)}$'s ($i=1,2$)  in (\ref{rep}) are then given by 
	\begin{align}
	&K_{21}\Phi_1^{(2)}(x,y) = 4p_1^{\Delta-2 d -4} \sum_{a,b} c^{(2)}_1(a,b,\vec{\Delta})\,x^a y^b \frac{\partial}{\partial x} \,F_4(\alpha(a.b) +1, \beta(a,b)+1; \gamma(a)-1, \gamma'(b) ; x, y)   \notag\\
	&= 4 p_1^{\Delta-2 d -4} \sum_{a,b} c^{(2)}_1(a,b,\vec{\Delta})\,x^a y^b \frac{(\alpha(a,b)+1)(\beta(a,b)+1)}{(\gamma(a)-1)}F_4(\alpha(a,b) +2, \beta(a,b)+2; \gamma(a), \gamma'(b); x, y) \notag\\
	&K_{31}\Phi_1^{(2)}(x,y)= 0\\[3ex]
	&K_{31}\Phi_2^{(2)}(x,y) = 4 p_1^{\Delta-2 d -4} \sum_{a,b} c^{(2)}_2(a,b,\vec{\Delta})\, x^a y^b \frac{\partial}{\partial y} \,F_4(\alpha(a,b) +1, \beta(a,b)+1; \gamma(a), \gamma'(b)-1 ; x, y) \notag\\
	&=4 p_1^{\Delta-2 d -4} \sum_{a,b} c^{(2)}_2(a,b,\vec{\Delta})\,  x^a y^b  \frac{(\alpha(a,b)+1)(\beta(a,b)+1)}{(\gamma'(b)-1)} F_4(\alpha(a,b) +2, \beta(a,b)+2; \gamma(a), \gamma'(b); x, y)
	\notag\\
	&K_{21} \Phi_2^{(2)}(x,y)=0,
	\end{align}
	where it is clear that the non-zero right-hand-side of both equations are proportional to the form factor $A_1$ given in (\ref{A1}). Once this particular solution is determined, Eq. \eqref{A1}, by comparison, gives the conditions on $c_1^{(2)}$ and $c_1^{(2)}$ as
	\begin{align}
	c_1^{(2)}(a,b,\vec{\Delta})&=\frac{\gamma(a)-1}{2(\alpha(a,b)+1)(\beta(a,b)+1)}\ c^{(1)}(a,b,\vec \Delta)\,,\\
	c_2^{(2)}(a,b,\vec{\Delta})&=\frac{\gamma'(b)-1}{2(\alpha(a,b)+1)(\beta(a,b)+1)}\ c^{(1)}(a,b,\vec \Delta)\,.
	\label{condc2}
	\end{align}
	We conclude that the general solution for $A_2$ in the $TJJ$ case (in which $\g(a)=\g'(b)$ ) is given by combining the solution of the homogeneous form of \eqref{A1} and the particular one given by \eqref{rep1} and \eqref{rep}, by choosing the constants appropriately using \eqref{condc2}. It is explicitly given by 
	\begin{align}
	A_2&= p_1^{\Delta-2 d - 2}\sum_{a b} x^a y^b\Bigg[c^{(2)}(a,b,\vec{\Delta})\,F_4(\alpha(a,b)+1, \beta(a,b)+1; \gamma(a), \gamma'(b); x, y)\notag\\
	&\hspace{1cm}+ \frac{(\gamma(a)-1)\,c^{(1)}(a,b,\vec \Delta)}{2(\alpha(a,b)+1)(\beta(a,b)+1)}\bigg(F_4(\alpha(a,b) +1, \beta(a,b)+1; \gamma(a)-1, \gamma'(b); x, y)\notag\\
	&\hspace{7cm}+ F_4(\alpha(a,b) +1, \beta(a,b)+1; \gamma(a), \gamma'(b)-1; x, y)\bigg)\Bigg],
	\end{align}
	 since $\g(a)=\g'(b)$. 
	The approach allows to solve for all the form factors $A_i$, if used sequentially. The derivation of the solution is discussed in \cite{Coriano:2018bbe}. A similar analysis allows to solve for other correlators, such as the $TTT$ \cite{Coriano:2018bsy}. Notice that a reduction of the number of constants generated from \eqref{Primary} is obtained by imposing some additionanal WIs which link special CWIs and canonical WI's. They have been called {\em secondary}  WIs in \cite{Bzowski:2013sza}. The solution of these additional lequations, which connect 3- to 2-point functions, can be solved by performing some specific limits on the momentum variables.
	
	\section{Dual conformal/conformal symmetry} 
	Dual conformal symmetry has been originally identified in the context of perturbative quantum field theory and holds for a special class of Feynman diagrams \cite{Drummond:2007aua}.  
	We recall that a dual conformal integral is a Feynman integral which, once rewritten in terms of some dual coordinates, under the action of $K^\kappa$ is modified by factors which depend only on the coordinates of the external points. The reformulation of the ordinary momentum integral in terms of such dual coordinates can be immediately worked out by drawing the associated dual diagram. 
	\begin{figure}[t]
		\centering
		\raisebox{-0.5\height}{\includegraphics[scale=0.3]{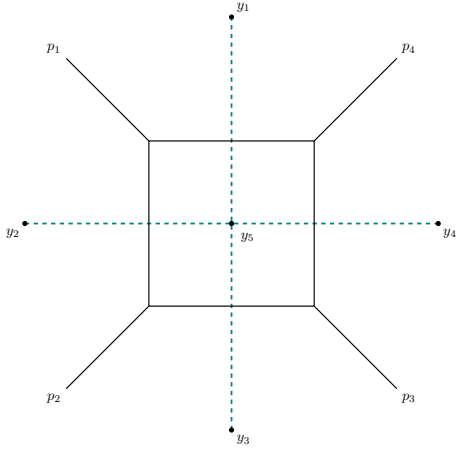}} \hspace{2cm}
		\caption{The Feynman diagram corresponding to the box topology with its dual. The momenta $p_i$ are all incoming. $y_i$ are dual variables in momentum space. \label{Fig1}}
	\end{figure}
	We illustrate this point in the case of the ordinary Feynman (box) diagram shown in Fig. \ref{Fig1}
	\begin{equation}
	\Phi_{Box}(p_1,p_2,p_3,p_4)=\int \frac{d^d k}{k^2 (k + p_1)^2 (k + p_1 + p_2)^2 ( k + p_1 + p_2 + p_3)^2},
	\end{equation}
	which is a function of 6 Lorentz invariants. As usual in particle theory, we use the six scalar variables $p_i^2$ $(i=1,2,3,4)$ and $s^2=(p_1+p_2)^2$ and $t^2=(p_1+p_3)^2$.
	We introduce dual variables $y_i$, redefining the momenta in terms of these
	\begin{equation}
	k=y_{51}, \qquad p_1=y_{12}, \qquad p_2=y_{23},\qquad p_3=y_{34},
	\label{map}
	\end{equation}
	with $y_{ij}=y_i-y_j$, and rewrite the integral in the form 
	\begin{equation}
	\label{box1}
	\Phi_{Box}(y_1,y_2,y_3,y_4)=\int \frac{d^d y_5}{y_{15}^2 y_{25}^2 y_{35}^2 y_{45}^2}.
	\end{equation}
	As already mentioned, the action of $K^\kappa$ is realized in the form $\mathcal{I} \cdot \mathcal{T}\cdot \mathcal{I}$, as a sequence of inversion, translation and inversion transformations, rather than as a differential action (by $K^\kappa$, as in \eqref{Scw}). We recall that under an inversion $(\mathcal{I})$ 
	\begin{equation}
	\mathcal{I} (d^d y_5)=d^d y_5 {(y_5^2)}^{-d}  \qquad \mathcal{I}(y^2_{ij})=\frac{y_{ij}^2}{y_i^2y_j^2}
	\end{equation}
	and in order to obtain an expression which is invariant under special conformal transformation, it is necessary to include a pre-factor in $\Phi_{Box}$, in the form  
	\begin{equation}
	s^2 t^2 \Phi_{Box}(p_1,p_2,p_3,p_4)= y_{13}^2 y_{24}^2  \Phi_{Box}(y_1,y_2,y_3,y_4),
	\end{equation}
	and then it is easy to check that under the action of $\mathcal{I}$ the integrand 
	\begin{equation}
	\mathcal{I} \left(\frac{d^d y_5 y_{13}^2 y_{2 4}^2}{y_{15}^2 y_{25}^2 y_{35}^2 y_{45}^2}\right)= 
	\left(\frac{d^d y_5   (y_5^2)^{4-d}  y_{13}^2 y_{2 4}^2}{y_{15}^2 y_{25}^2 y_{35}^2 y_{45}^2}\right)
	\end{equation}
	is invariant if $d=4$. Obviously, the invariance under the complete action $\mathcal{I T I}$ is ensured. It is easily checked that the integrand is also scale invariant. It is then clear that the expression of the box diagram in $d=4$ can only be of the form 
	\begin{equation}
	\label{ans1}
	\Phi_{Box}=\frac{1}{y_{13}^2 y_{24}^2}F\big(u(y_i),v(y_i)\big)
	\end{equation}
	with $u$ and $v$ given by 
	\begin{equation}
	\label{uv2}
	u(y_i)=\frac{y_{12}^2 y_{34}^2}{y_{13}^2 y_{24}^2} \qquad v(y_i)=\frac{y_{23}^2 y_{41}^2}{y_{13}^2 y_{24}^2},
	\end{equation}
which can be expressed directly in terms of the original momentum invariants, due to the relations \eqref{map}.
	Notice that, by construction, $u$ and $v$ satisfy the first order equations in the $y$ variables
	\begin{equation}
	\label{firstp}
	\begin{split}
	K_0^\kappa(y)\, u(y_i) &\equiv\sum_{j=1}^{4} \left(- y_j^2\frac{\partial}{\partial y_j^\kappa}+ 2 y_j^\kappa y_j^\alpha \frac{\partial}
	{\partial y_j^\alpha} \right)  u(y_i) =0\\
	K_0^\kappa(y)\, v(y_i) &\equiv\sum_{j=1}^{4} \left(- y_j^2\frac{\partial}{\partial y_j^\kappa}+ 2 y_j^\kappa y_j^\alpha \frac{\partial}
	{\partial y_j^\alpha} \right)  v(y_i) =0.
	\end{split}
	\end{equation}
	Dual conformal Feynman diagrams, such as the one discussed above, satisfy in the $y$ variables \eqref{map} CWI's as in ordinary coordinate space. The $y$ variables 
	are, however, variables of momentum space, for being linear combinations of the fundamental momenta $p_i$, obtained from the Fourier transform of the original correlator in coordinate 
	space.   \\
	It is then interesting to search for solutions of the 4-point functions which are at the same time 
	conformal in the coordinate variables $(x_i)$, as well in the auxiliary variables $y_i$ introduced by the mapping \eqref{map}. \\
	This implies that a scalar 4-point function, once written in momentum space and re-expressed in terms of variables $y_i$, has necessarily to take a form similar to \eqref{ans1}, for being conformal in such variables as well. The simultaneous conditions of conformal and dual conformal invariance are then satisfied if a scalar correlator takes the form \eqref{ans1} and, at the same time, satisfies the ordinary CWI in the ordinary momentum variables $p_i$, once the $y$ variables are re-expressed in terms of the momenta using  \eqref{map}. We have shown in \cite{Maglio:2019grh} that the form taken by $F(u,v)$ is unique. We are going to illustrate the construction of such solutions.
	
	\subsection{The CWIs for scalar 4-point correlators} 
	To illustrate the approach in a realistic case, let's consider a generic scalar 4-point function in momentum space
	
	\begin{equation}
	\braket{O(\mathbf{p}_1)\,O(\mathbf{p}_2)\,O(\mathbf{p}_3)\,O(\bar{\mathbf{p}}_4)}=\Phi(p_1,p_2,p_3,p_4,s,t),\label{invariant}
	\end{equation}
	with the definitions of the invariants and Mandelstam variables as
	\begin{equation}\label{defst}
	p_i=|\mathbf{p}_i|,\quad s=|\mathbf{p}_1+\mathbf{p_2}|, \quad t=|\mathbf{p}_2+\mathbf{p}_3|.
	\end{equation} 
	This correlation function, to be conformal invariant, has to verify the dilatation Ward Identity
	\begin{equation}
	\left[\sum_{i=1}^4\D_i-3d-\sum_{i = 1}^3p_i^{\mu}\frac{\partial}{\partial p_i^\mu}\right]\braket{O(\mathbf{p}_1)\,O(\mathbf{p}_2)\,O(\mathbf{p}_3)\,O(\bar{\mathbf{p}}_4)}=0
	\end{equation}
	and the special conformal Ward Identities
	\begin{equation}
	\sum_{i=1}^3\left[2(\D_i-d)\frac{\partial}{\partial p_{i\,\k}}-2p_i^\alpha\frac{\partial^2}{\partial p_i^\alpha\partial p_i^\kappa}+p_i^\kappa\frac{\partial^2}{\partial p_i^\alpha\partial p_{i\,\alpha}}\right]\braket{O(\mathbf{p}_1)\,O(\mathbf{p}_2)\,O(\mathbf{p}_3)\,O(\bar{\mathbf{p}}_4)}=0,
	\end{equation}
	which are conditions generated by the conformal symmetry in the ordinary variables $x_i$, parameterizing the correlator in coordinate space.
	By turning to momentum space and using the chain rules 
	\begin{align}
	\frac{\partial}{\partial p_{1\,\mu}}&=\frac{p_1^\mu}{p_1}\frac{\partial}{\partial p_1}-\frac{\bar{p}_4^\mu}{p_4}\frac{\partial}{\partial p_4}+\frac{p_1^\mu+p_2^\mu}{s}\frac{\partial}{\partial s}\\
	\frac{\partial}{\partial p_{2\,\mu}}&=\frac{p_2^\mu}{p_2}\frac{\partial}{\partial p_2}-\frac{\bar{p}_4^\mu}{p_4}\frac{\partial}{\partial p_4}+\frac{p_1^\mu+p_2^\mu}{s}\frac{\partial}{\partial s}+\frac{p_2^\mu+p_3^\mu}{t}\frac{\partial}{\partial t}
	\end{align}
	and other similar ones, one derives the two equations
	\begin{align}
	C_2&=\bigg\{\frac{\partial^2}{\partial p_2^2}+\frac{(d-2\D_2+1)}{p_2}\frac{\partial}{\partial p_2}-\frac{\partial^2}{\partial p_4^2}-\frac{(d-2\D_4+1)}{p_4}\frac{\partial}{\partial p_4}\notag\\
	&\qquad+\frac{1}{s}\frac{\partial}{\partial s}\left(p_1\frac{\partial}{\partial p_1}+p_2\frac{\partial}{\partial p_2}-p_3\frac{\partial}{\partial p_3}-p_4\frac{\partial}{\partial p_4}\right)+\frac{(\D_3+\D_4-\D_1-\D_2)}{s}\frac{\partial}{\partial s}\notag\\
	&\qquad+\frac{1}{t}\frac{\partial}{\partial t}\left(p_2\frac{\partial}{\partial p_2}+p_3\frac{\partial}{\partial p_3}-p_1\frac{\partial}{\partial p_1}-p_4\frac{\partial}{\partial p_4}\right)+\frac{(\D_1+\D_4-\D_2-\D_3)}{t}\frac{\partial}{\partial t}\notag\\[1.2ex]
	&\qquad+\frac{(p_2^2-p_4^2)}{st}\frac{\partial^2}{\partial s\partial t}\bigg\}\,\Phi(p_1,p_2,p_3,p_4,s,t)=0
	\label{C2}
	\end{align}
	\begin{align}
	C_{13}&=\bigg\{\frac{\partial^2}{\partial p_1^2}+\frac{(d-2\D_1+1)}{p_1}\frac{\partial}{\partial p_1}-\frac{\partial^2}{\partial p_3^2}-\frac{(d-2\D_3+1)}{p_3}\frac{\partial}{\partial p_3}\notag\\
	&\qquad+\frac{1}{s}\frac{\partial}{\partial s}\left(p_1\frac{\partial}{\partial p_1}+p_2\frac{\partial}{\partial p_2}-p_3\frac{\partial}{\partial p_3}-p_4\frac{\partial}{\partial p_4}\right)+\frac{(\D_3+\D_4-\D_1-\D_2)}{s}\frac{\partial}{\partial s}\notag\\
	&\qquad+\frac{1}{t}\frac{\partial}{\partial t}\left(p_1\frac{\partial}{\partial p_1}+p_4\frac{\partial}{\partial p_4}-p_2\frac{\partial}{\partial p_2}-p_3\frac{\partial}{\partial p_3}\right)+\frac{(\D_2+\D_3-\D_1-\D_4)}{t}\frac{\partial}{\partial t}\notag\\[1.2ex]
	&\qquad+\frac{(p_1^2-p_3^2)}{st}\frac{\partial^2}{\partial s\partial t}\bigg\}\,\Phi(p_1,p_2,p_3,p_4,s,t)=0,\label{Eq2}
	\end{align}
	together with a third one, that we omit, which takes a similar form. A detailed discussion of such systems of equations has been presented in \cite{Maglio:2019grh}. 
	Clearly, these equations do not define ,in general, a hypergeometric system. However, once we require that the solutions of these equations are also invariant in the momentum variables $y_{ij}$, where the $y_{ij}$ are treated as ordinary coordinates $x_i$, we generate a hypergeometric system of equations also for such 4-point functions.\\
As we have already mentioned,	the strategy to solve the equations is to start with an expression of these correlators of the form \eqref{ans} and impose on them the conditions of conformal invariance. Clearly, both these conditions are only formulated in momentum space.
	
	\subsection{DCC solutions}
	To identify the dual conformal/conformal solutions, we choose  the ansatz
	\begin{equation}
	\Phi(p_i,s,t)=\big(s^2t^2\big)^{n_s}\,F(x,y),\label{ansatz}
	\end{equation}
	where $n_s$ is a coefficient (scaling factor of the ansatz) that we will fix below by the dilatation WI, and the variables $x$ and $y$ are defined by the quartic ratios
	\begin{equation}
	x=\frac{p_1^2\,p_3^2}{s^2\,t^2},\qquad y=\frac{p_2^2\,p_4^2}{s^2\,t^2}.
	\label{xandy}
	\end{equation}
	Such quartic ratios are nothing else but the two variables $u$ and $v$ of \eqref{uv2} re-expressed in momentum space.\\
	By inserting the ansatz \eqref{ansatz} into the dilatation Ward Identities, and turning to the new variables $x$ and $y$, after some manipulations we obtain from the dilatation WI the constraint \begin{align}
	\label{dil1}
	&\bigg[(\D_t-3d)-\sum_{i=1}^4p_i\frac{\partial}{\partial p_i}-s\frac{\partial}{\partial s}-t\frac{\partial}{\partial t}\bigg]\big(s^2t^2\big)^{c}\,F(x,y),\notag\\
	&=\big(s^2t^2\big)^{n_s}\big[(\D_t-3d)-4 n_s\big] \,F(x,y)=0
	\end{align}
	which determines $n_s=(\D_t-3d)/4$, giving
	\begin{equation}
	\Phi(p_i,s,t)=\big(s^2t^2\big)^{(\D_t-3d)/4}\,F(x,y).\label{ansatz2}
	\end{equation}
	We will be using this specific form of the solution in two of three special CWIs of the form  $C_2$ and $C_{13}$. The functional form of $F(x,y)$ will then be constrained further. 
	
	In order to determine the conditions on $F(x,y)$ from \eqref{C2} and \eqref{Eq2},  we re-express these two equations in terms of $x$ and $y$ using several identities. In particular we will use the relations 
	\begin{equation}
	\frac{\partial^2}{\partial s\partial t}F(x,y)=\frac{4}{st}\big[\big(
	x\,\partial_x +y\partial_y\big)F+\big(x^2\partial_{xx}+2xy\,\partial_{xy}+y^2\partial_{yy}\big)F\big],
	\label{onew}
	\end{equation}
	together with
	\begin{equation}
	\begin{split}
	\left(p_1\frac{\partial}{\partial p_1}+p_2\frac{\partial}{\partial p_2}-p_3\frac{\partial}{\partial p_3}-p_4\frac{\partial}{\partial p_4}\right)F(x,y)&=\left(2x\,\partial_x+2y\,\partial_y -2x\,\partial_x-2y\,\partial_y\right)F(x,y)=0,\\[1.5ex]
	\left(p_1\frac{\partial}{\partial p_1}+p_4\frac{\partial}{\partial p_4}-p_3\frac{\partial}{\partial p_3}-p_2\frac{\partial}{\partial p_2}\right)F(x,y)&=\left(2x\,\partial_x+2y\,\partial_y -2x\,\partial_x-2y\,\partial_y\right)F(x,y)=0.
	\end{split}
	\label{twow}
	\end{equation}
	Both relations can be worked out after some lengthy computations. \\
	We start investigating the solutions of these equations by assuming, for example, that the scaling dimensions of all the fields $\phi_i$ are  equal $\D_1=\D_2=\D_3=\D_4=\D$.\\
	Using \eqref{onew} and \eqref{twow}, we write the first equation \eqref{C2} associated to $C_2$ in the new variable $x$ and $y$ as
	\begin{align}
	&C_2= 4\big(p_2^2-p_4^2\big)(s^2)^{n_s-1}(t^2)^{n_s-1}\notag\\
	&\times\bigg[y(1-y)\partial_{yy} -2x\,y\,\partial_{xy}-x^2\partial_{xx}-(1-2n_s)x\,\partial_x+\left(1-\D+\frac{d}{2}-y(1-2n_s)\right)\,\partial_y-n_s^2\bigg] F(x,y)=0\label{first}
	\end{align}
	and the second one \eqref{Eq2} associated to $C_{13}$ as
	\begin{align}
	&C_{13}=4\big(p_1^2-p_3^2\big)(s^2)^{n_s-1}(t^2)^{n_s-1}\notag\\
	&\times\bigg[x(1-x)\partial_{xx} -2x\,y\,\partial_{xy}-y^2\partial_{yy}-(1-2n_s)y\,\partial_y+\left(1-\D+\frac{d}{2}-x(1-2n_s)\right)\,\partial_x-n_s^2\bigg] F(x,y)=0\label{second}
	\end{align}
	where we recall that $n_s$ is the scaling under dilatations, now given by
	\begin{equation}
	\label{scaled}
	n_s=\D-\frac{3d}{4}
	\end{equation}
	since $\D_t=4\, \D$. \\
	By inspection, one easily verifies that \eqref{first} and \eqref{second} define a hypergeometric system of two equations whose solutions can be expressed as linear combinations of 4 Appell functions of two variables $F_4$, as in the case of 3-point functions discussed before. The general solution of such system is expressed as
	\begin{align}
	\Phi(p_i,s,t) &=\big(s^2t^2\big)^{(\D_t-3d)/4}\, F(x,y) \notag\\
	F(x,y)&= \sum_{a,b} c(a,b,\vec\Delta_t) x^a y^b F_4\left(\a(a,b),\b(a,b),\g(a),\g'(b);x, y\right),\label{solution}
	\end{align}
	with $\vec\Delta_t=\Delta(1,1,1,1)$, if we choose the operators of equal scalings.  Notice that the solution is similar to that of the 3-point functions given in the previous section.\\
	The general solution \eqref{solution} has been written as a linear superposition of these with independent constants $c(a,b)$, labelled by the exponents $a,b$ 
	\begin{align}
	&a=0,\,\D-\frac{d}{2},&&b=0,\,\D-\frac{d}{2},\label{FuchsianPoint}
	\end{align}
	which fix the dependence of the $F_4$  
	\begin{align}
	\label{s2}
	&\a(a,b)=\frac{3}{4}d-\D+a+b,&&\b(a,b)=\frac{3}{4}d-\D+a+b,\notag\\
	&\g(a)=\frac{d}{2}-\D+1+2a,&&\g'(b)=\frac{d}{2}-\D+1+2b.
	\end{align}
	The proof that the ansatz \eqref{ansatz} satisfies the CWIs as given by $C_2$ and $C_{13}$  we use the following identities for the Appell hypergeometric function 
	\begin{align}
	\partial_x\,F_4(a,b,c_1,c_2;x,y)&=\frac{a\,b}{c_1}\,F_4(a+1,b+1,c_1+1,c_2;x,y)\\
	\partial_y\,F_4(a,b,c_1,c_2;x,y)&=\frac{a\,b}{c_2}\,F_4(a+1,b+1,c_1,c_2+1;x,y)
	\end{align}
	\begin{equation}
	x\,\partial_x\, F_4(a,b,c_1,c_2;x,y)= (c_1-1)\big[F_4(a,b,c_1-1,c_2;x,y)-F_4(a,b,c_1,c_2;x,y)\big].
	\end{equation} 
	We can use these identities to derive the relations 
	\begin{align}
	x\,\partial_{xx}\, F_4(a,b,c_1,c_2;x,y)&=(c_1-1) \partial_x\,F_4(a,b,c_1-1,c_2;x,y)-c_1\,\partial_x\,F_4(a,b,c_1,c_2;x,y)\notag\\
	&=a\,b\big[F_4(a+1,b+1,c_1,c_2;x,y)-F_4(a+1,b+1,c_1+1,c_2;x,y)\big]
	\end{align}
	with an analogous expression obtained for the $y$ variable. The intermediate steps are rather involved and can be found in \cite{Maglio:2019grh}. The final conclusion is that the original ansatz \eqref{ansatz} indeed satisfies the CWI's in momentum space if the function $F(x,y)$ is a hypergeometric function of the new variables x and y given in \eqref{xandy}. The solution takes the form 
	\begin{equation}
	\resizebox{1\hsize}{!}{$
		\begin{aligned}
		\Phi(p_1,p_2,p_3,p_4,s,t)&=c_1\bigg\{\left(p_1^2\,p_3^2\right)^{\D-\frac{3}{4}d}\bigg[F_4\left(\frac{d}{4}\,,\,\frac{3}{4}d-\D\,,\,1\,,\,\frac{d}{2}-\D+1\,;\frac{s^2t^2}{p_1^2p_3^2}\,,\,\frac{p_2^2p_4^2}{p_1^2p_3^2}\right)\\[1.2ex]
		&\hspace{-0.3cm}+\frac{\Gamma\left(\D-\frac{d}{4}\right)\Gamma\left(1+\D-\frac{3}{4}d\right)\Gamma\left(1-\D+\frac{d}{2}\right)}{\Gamma\left(\D-\frac{3}{4}d\right)\Gamma\left(1-\D+\frac{3}{4}d\right)\Gamma\left(1+\D-\frac{d}{2}\right)}\left(\frac{p_2^2p_4^2}{p_1^2p_3^2}\right)^{\D-\frac{d}{2}} F_4\left(\D-\frac{d}{4}\,,\,\frac{d}{4}\, ,\,1\,,\,1-\frac{d}{2}+\D\,;\frac{s^2t^2}{p_1^2p_3^2}\,,\,\frac{p_2^2p_4^2}{p_1^2p_3^2}\right)\bigg]\\[1.2ex]
		&+\left(p_2^2\,p_3^2\right)^{\D-\frac{3}{4}d}\bigg[F_4\left(\frac{d}{4}\,,\,\frac{3}{4}d-\D\, ,\,1\,,\,\frac{d}{2}-\D+1\,;\frac{s^2u^2}{p_2^2p_3^2}\,,\,\frac{p_1^2p_4^2}{p_2^2p_3^2}\right)\\[1.2ex]
		&\hspace{-0.3cm}+\frac{\Gamma\left(\D-\frac{d}{4}\right)\Gamma\left(1+\D-\frac{3}{4}d\right)\Gamma\left(1-\D+\frac{d}{2}\right)}{\Gamma\left(\D-\frac{3}{4}d\right)\Gamma\left(1-\D+\frac{3}{4}d\right)\Gamma\left(1+\D-\frac{d}{2}\right)} \left(\frac{p_1^2p_4^2}{p_2^2p_3^2}\right)^{\D-\frac{d}{2}} F_4\left(\D-\frac{d}{4}\,,\,\frac{d}{4}\, ,\,1\,,\,1-\frac{d}{2}+\D\,;\frac{s^2u^2}{p_2^2p_3^2}\,,\,\frac{p_1^2p_4^2}{p_2^2p_3^2}\right)\bigg]\\[1.2ex]
		&+\left(p_1^2\,p_2^2\right)^{\D-\frac{3}{4}d}\bigg[F_4\left(\frac{d}{4}\,,\,\frac{3}{4}d-\D\, ,\,1\,,\,\frac{d}{2}-\D+1\,;\frac{u^2t^2}{p_1^2p_2^2}\,,\,\frac{p_3^2p_4^2}{p_1^2p_2^2}\right)\\[1.2ex]
		&\hspace{-0.3cm}+\frac{\Gamma\left(\D-\frac{d}{4}\right)\Gamma\left(1+\D-\frac{3}{4}d\right)\Gamma\left(1-\D+\frac{d}{2}\right)}{\Gamma\left(\D-\frac{3}{4}d\right)\Gamma\left(1-\D+\frac{3}{4}d\right)\Gamma\left(1+\D-\frac{d}{2}\right)}\left(\frac{p_3^2p_4^2}{p_1^2p_2^2}\right)^{\D-\frac{d}{2}} F_4\left(\D-\frac{d}{4}\,,\,\frac{d}{4}\, ,\,1\,,\,1-\frac{d}{2}+\D\,;\frac{u^2t^2}{p_1^2p_2^2}\,,\,\frac{p_3^2p_4^2},{p_1^2p_2^2}\right)\bigg]\bigg\},\label{finalsolution}
		\end{aligned}$}
	\end{equation}
	which can be shown to be symmetric under all the possible permutations of the momenta $(p_1,\,p_2,\,p_3,\,p_4)$ and $c_1$ is an overall constant. As shown in \cite{Maglio:2019grh} such solution is unique. This property is guaranteed by the absence of unphysical singularities in the domain of convergence of the solution found.\\
	Also in this case we can reformulate such solution in terms of 3K integrals using the expression 
	\begin{align}
	&I_{\frac{d}{2}-1\{\Delta-\frac{d}{2},\Delta-\frac{d}{2},0\}}(p_1p_3,p_2p_4,s,t)=\notag\\
	&\qquad=\,(p_1p_3)^{\Delta-\frac{d}{2}}(p_2p_4)^{\Delta-\frac{d}{2}}\int_{0}^\infty\,dx\,x^{\frac{d}{2}-1}\,K_{\Delta-\frac{d}{2}}(p_1p_3\,x)\,K_{\Delta-\frac{d}{2}}(p_2p_4\,x)\,K_{0}(st\,x),\label{Sol}
	\end{align}
	which is close in form to the result obtained for 3-point functions, but now expressed in terms of quartic ratios of momenta.
	Using  \eqref{der} one can derive several relations, 
such as 
\begin{align}
\frac{\partial^2}{\partial p_1^2}I_{\a\{\b_1,\b_2,\b_3\}}&=-\,p_3^2\,I_{\a+1\{\b_1-1,\b_2,\b_3\}}+p_1^2\,p_3^4\,\,I_{\a+2\{\b_1-2,\b_2,\b_3\}}
\end{align}
which generate identities such as 
\begin{align}
p_1^2\,p_3^2\,I_{\a+2\{\b_1-2,\b_2,\b_3\}}&=I_{\a+2\{\b_1,\b_2,\b_3\}}-2(\b_1-1)\,I_{\a+1\{\b_1-1,\b_2,\b_3\}}.
\end{align}
One can show that the $I$ integrals satisfy the differential equations
\begin{align}
\frac{1}{s}\frac{\partial}{\partial s}\left(p_1\frac{\partial}{\partial p_1}+p_2\frac{\partial}{\partial p_2}-p_3\frac{\partial}{\partial p_3}-p_4\frac{\partial}{\partial p_4}\right)I_{\a\{\b_1,\b_2,\b_3\}}&=0\\
\frac{1}{t}\frac{\partial}{\partial t}\left(p_1\frac{\partial}{\partial p_1}+p_4\frac{\partial}{\partial p_4}-p_2\frac{\partial}{\partial p_2}-p_3\frac{\partial}{\partial p_3}\right)I_{\a\{\b_1,\b_2,\b_3\}}&=0.
\end{align} 

In the case $\D_1=\D_3=\D_x$ and $\D_2=\D_4=\D_y$, the special CWI's can be written as
\begin{align}
\braket{O(\mathbf{p}_1)\,O(\mathbf{p}_2)\,O(\mathbf{p}_3)\,O(\bar{\mathbf{p}}_4)}&=\,\bar{\bar{\a}} \,I_{\frac{d}{2}-1\left\{\D_x-\frac{d}{2},\D_y-\frac{d}{2},0\right\}}(p_1\, p_3;p_2\,p_4; s\,t),
\end{align}
with $\bar{\bar{\a}}$ an arbitrary constant, in agreement with the solution found for the three-point function. 

\section{Lauricella functions }
We now come to the last part of this overview, illustrating the appearance of another type of functions of hypergeometric type in this class of equations. \\
Lauricella systems of equations are generated if we look for asymptotic solutions of the CWIs characterised by large $s$ and $t$ invariants, under the condition that $p_i^2 \ll s^2, t^2$, recalling the definition of these variables in \eqref{defst}. We remind that in a 2-to-2 process 
described by 4-point functions, a scattering at fixed angle is characterised by such invariants with the ratio $-t/s$ held fixed. We are then allowed to take both $s$ and $t$ large, keeping their ratio fixed. In this limit one can show that the CWIs are approximately separable and the solutions satisfy a factorized form in which we separate the dependence of $\Phi$ on the invariants $p_i^2$ from the pair $s$ and $t$.\\
It is also possible to show that a Lauricella system describes a particular solution of the special CWIs of the form $C_2$ and $C_{13}$. The analysis has been presented in \cite{Maglio:2019grh} and \cite{Coriano:2019nkw}, to which we refer for further details. \\
	The equations in this case take the form
	\begin{equation}
	\textup{K}_{14}\phi=0,\qquad \textup{K}_{24}\phi=0,\qquad \textup{K}_{34}\phi=0\label{CWILaur}.
	\end{equation}
	This operator can be rearranged by introducing a change of variables of the form $(p_1^2,p_2^2,p_3^2,p_4^2)$ to $(x,y,z,p_4^2)$ where 
	\begin{equation}
	x=\frac{p_1^2}{p_4^2},\quad y=\frac{p_2^2}{p_4^2},\quad z=\frac{p_3^2}{p_4^2}
	\end{equation}
	are the dimensionless ratios $x, y$ and $z$, not to be confused with coordinate points in a three dimensional space. The ansatz for the solution can be taken of the form
	\begin{equation}
	\phi(p_1,p_2,p_3,p_4)=(p_4^2)^{n_s}\,x^a\,y^b\,z^c\,F(x,y,z),
	\end{equation}
	which is consistent with the corresponding dilatation WI for 4-point functions if
	
	\begin{equation}
	n_s=\frac{\D_t}{2}-\frac{3d}{2}.
	\end{equation}
	With this ansatz, equations \eqref{CWILaur} take the form
	\begin{align}
	\textup{K}_{14}\phi=&4p_4^{\D_t-3d-2}\,x^a\,y^b\,z^c\,\bigg[(1-x)x\frac{\partial^2}{\partial x^2}-2x\,y\frac{\partial^2}{\partial x\partial y}-y^2\frac{\partial^2}{\partial y^2}-2x\,z\frac{\partial^2}{\partial x\partial z}-z^2\frac{\partial^2}{\partial z^2}-2y\,z\frac{\partial^2}{\partial y\partial z}\notag\\
	&\hspace{2cm}+(Ax+\gamma)\frac{\partial}{\partial x}+Ay\frac{\partial}{\partial y}+Az\frac{\partial }{\partial z}+\left(E+\frac{G}{x}\right)\bigg]F(x,y,z)=0
	\end{align}
	with
	\begin{subequations}
		\begin{align}
		A&=\D_1+\D_2+\D_3-\frac{5}{2}d-2(a+b+c)-1\\
		E&=-\frac{1}{4}\big(3d-\D_t+2(a+b+c)\big)\big(2d+2\D_4-\D_t+2(a+b+c)\big)\\
		G&=\frac{a}{2}\,\left(d-2\D_1+2a\right)\\
		\g&=\frac{d}{2}-\D_1+2a+1.
		\end{align}
	\end{subequations}
	Similar constraints are obtained from the equation $\textup{K}_{34}\phi=0$ that can be written as
	\begin{align}
	\textup{K}_{24}\phi=&4p_4^{\D_t-3d-2}\,x^a\,y^b\,z^c\,\bigg[-x^2\frac{\partial^2}{\partial x^2}-2x\,y\frac{\partial^2}{\partial x\partial y}+(1-y)y\frac{\partial^2}{\partial y^2}-2x\,z\frac{\partial^2}{\partial x\partial z}-z^2\frac{\partial^2}{\partial z^2}-2y\,z\frac{\partial^2}{\partial y\partial z}\notag\\
	&\hspace{2cm}+A'x\frac{\partial}{\partial x}+(A'y+\g')\frac{\partial}{\partial y}+A'z\frac{\partial }{\partial z}+\left(E'+\frac{G'}{x}\right)\bigg]F(x,y,z)=0,
	\end{align}
	with
	\begin{subequations}
		\begin{align}
		A'&=\D_1+\D_2+\D_3-\frac{5}{2}d-2(a+b+c)-1\\
		E'&=-\frac{1}{4}\big(3d-\D_t+2(a+b+c)\big)\big(2d+2\D_4-\D_t+2(a+b+c)\big)\\
		G'&=\frac{b}{2}\,\left(d-2\D_2+2b\right)\\
		\g'&=\frac{d}{2}-\D_2+2b+1,
		\end{align}
	\end{subequations}
	and 
	\begin{align}
	\textup{K}_{34}\phi=&4p_4^{\D_t-3d-2}\,x^a\,y^b\,z^c\,\bigg[-x^2\frac{\partial^2}{\partial x^2}-2x\,y\frac{\partial^2}{\partial x\partial y}-y^2\frac{\partial^2}{\partial y^2}-2x\,z\frac{\partial^2}{\partial x\partial z}+(1-z)z\frac{\partial^2}{\partial z^2}-2y\,z\frac{\partial^2}{\partial y\partial z}\notag\\
	&\hspace{2cm}+A''x\frac{\partial}{\partial x}+A''y\frac{\partial}{\partial y}+(A''z+\g'')\frac{\partial }{\partial z}+\left(E''+\frac{G''}{x}\right)\bigg]F(x,y,z)=0,
	\end{align}
	with
	\begin{subequations}
		\begin{align}
		A''&=\D_1+\D_2+\D_3-\frac{5}{2}d-2(a+b+c)-1\\
		E''&=-\frac{1}{4}\big(3d-\D_t+2(a+b+c)\big)\big(2d+2\D_4-\D_t+2(a+b+c)\big)\\
		G''&=\frac{c}{2}\,\left(d-2\D_3+2c\right)\\
		\g''&=\frac{d}{2}-\D_3+2c+1.
		\end{align}
	\end{subequations}
	Also in this case, as for 3-poin functions, in order to perform the reduction to the hypergeometric form of the equations, we need to set $G=0$, $G'=0$ and $G''=0$, which imply that the Fuchsian points $a,b,c$ take the values
	\begin{subequations}
		\begin{align}
		a&=0,\,\D_1-\frac{d}{2}\\
		b&=0,\,\D_2-\frac{d}{2}\\
		c&=0,\,\D_3-\frac{d}{2}.
		\end{align}
	\end{subequations}
	We find also that $E=E'=E''=-\a(a,b,c)\,\b(a,b,c)$ where
	\begin{align}
	\a(a,b,c)&=d+\D_4-\frac{\D_t}{2}+a+b+c\notag\\
	\b(a,b,c)&=\frac{3d}{2}-\frac{\D_t}{2}+a+b+c
	\end{align}
	as well as $A=A'=A''=-(\a(a,b,c)+\b(a,b,c)+1)$, indeed
	\begin{align}
	A=A'=A''&=-(\a(a,b,c)+\b(a,b,c)+1)=\D_1+\D_2+\D_3-\frac{5}{2}d-2(a+b+c)-1
	\end{align}
	and finally
	\begin{equation}
	\g(a)=\frac{d}{2}-\Delta_1+2a+1\,,\qquad\g'(b)=\frac{d}{2}-\Delta_2+2b+1\,,\qquad\g''(c)=\frac{d}{2}-\Delta_3+2c+1.
	\end{equation}
	With this redefinition of the coefficients, the equations are then expressed in the form 
	\cite{Maglio:2019grh}
	\begin{equation}
	\resizebox{1\hsize}{!}{$
		\left\{
		\begin{matrix}
		&x_j(1-x_j)\frac{\partial^2F}{\partial x_j^2}+\hspace{-1cm}\sum\limits_{\substack{\hspace{1.3cm}s\ne j\ \text{for}\ r=j}}\hspace{-1.1cm}x_r\hspace{0.2cm}\sum x_s\hspace{0.5ex}\frac{\partial^2F}{\partial x_r\partial x_s}+\left[\g_j-(\a+\b+1)x_j\right]\frac{\partial F}{\partial x_j}-(\a+\b+1)\sum\limits_{k\ne j}\,x_k\frac{\partial F}{\partial x_k}-\a\,\b\,F=0\\[3ex]
		& (j=1,2,3)
		\end{matrix}\right.\label{systemLauricella}$}
	\end{equation}
	having redefined $\g_1=\g$, $\ \g_2=\g'$ and $\g_3=\g''$ and $x_1=x$, $x_2=y$ and $x_3=z$. 
	This system of equations allows solutions in the form of the Lauricella hypergeometric function $F_C$ of three variables, which are defined by the series 
	\begin{equation}
	F_C(\a,\b,\g,\g',\g'',x,y,z)=\sum\limits_{m_1,m_2,m_3}^\infty\,\frac{(\a)_{m_1+m_2+m_3}(\b)_{m_1+m_2+m_3}}{(\g)_{m_1}(\g')_{m_2}(\g'')_{m_3}m_1!\,m_2!\,m_3!}x^{m_1}y^{m_2}z^{m_3},
	\end{equation}
	where the Pochhammer symbol $(\l)_{k}$ with an arbitrary $\l$ and $k$ a positive integer not equal to zero, was previously defined in \eqref{Pochh}. The convergence region of this series is defined by the condition
	\begin{equation}
	\left|\sqrt{x}\right|+\left|\sqrt{y}\right|+\left|\sqrt{z}\right|<1.
	\end{equation}
	The function $F_C$ is the generalization of the Appell $F_4$ to the case of three variables.
	The system of equations \eqref{systemLauricella} admits 8 independent solutions given by    \begin{align}
	&S_1(\a,\b,\g,\g',\g'',x,y,z)=F_C\big(\a,\b,\g,\g',\g'',x,y,z\big)\notag,\\
	&S_2(\a,\b,\g,\g',\g'',x,y,z)=x^{1-\g}\,F_C\big(\a-\g+1,\b-\g+1,2-\g,\g',\g'',x,y,z\big)\notag\,,\\
	&S_3(\a,\b,\g,\g',\g'',x,y,z)= y^{1-\g'}\,F_C\big(\a-\g'+1,\b-\g'+1,\g,2-\g',\g'',x,y,z\big)\notag\,,\\
	&S_4(\a,\b,\g,\g',\g'',x,y,z)=z^{1-\g''}\,F_C\big(\a-\g''+1,\b-\g''+1,\g,\g',2-\g'',x,y,z\big)\notag,\,\\
	&S_5(\a,\b,\g,\g',\g'',x,y,z)=x^{1-\g}y^{1-\g'}\,F_C\big(\a-\g-\g'+2,\b-\g-\g'+2,2-\g,2-\g',\g'',x,y,z\big)\,,\notag\\
	&S_6(\a,\b,\g,\g',\g'',x,y,z)=x^{1-\g}z^{1-\g''}\,F_C\big(\a-\g-\g''+2,\b-\g-\g''+2,2-\g,\g',2-\g'',x,y,z\big)\notag\,,\\
	&S_7(\a,\b,\g,\g',\g'',x,y,z)=y^{1-\g'}z^{1-\g''}\,F_C\big(\a-\g'-\g''+2,\b-\g'-\g''+2,\g,2-\g',2-\g'',x,y,z\big)\notag\,,\\
	&S_8(\a,\b,\g,\g',\g'',x,y,z)=x^{1-\g}y^{1-\g'}z^{1-\g''}\notag\\
	&\hspace{4cm}\times\,F_C\big(\a-\g-\g'-\g''+2,\b-\g-\g'-\g''+2,2-\g,2-\g',2-\g'',x,y,z\big)\,,\label{oneeq}
	\end{align} 
	having defined
	\begin{align}
	\a&\equiv\a(0,0,0)=d+\D_4-\frac{\D_t}{2}\notag\\    
	\b&\equiv\b(0,0,0)=\frac{3d}{2}-\frac{\D_t}{2}\notag\\
	\g&\equiv\g(0)=\frac{d}{2}-\Delta_1+1\notag\\
	\g'&\equiv\g'(0)=\frac{d}{2}-\Delta_2+1\notag\\
	\g''&\equiv\g''(0)=\frac{d}{2}-\Delta_3+1.
	\end{align}
	The most general solution of the system is obtained by taking linear combinations of such fundamental solutions with arbitrary constants. In this case, as in the cases discussed above for 3-point functions, one needs to impose the symmetry of the solution under exchanges of the momenta and of the scaling dimensions $\Delta_i$. This is most easily accomplished by establishing a link between the Lauricella function $F_C$ and some parametric integrals of 4 Bessel functions, which have been introduced in \cite{Maglio:2019grh}. 
	
\subsection{Lauricella's as 4-K integrals}
The introduction of parametric representations of the solutions of the CWI's in terms of 3K integrals \eqref{trekappa} 
in \cite{Bzowski:2013sza} has as its advantage the possibility of implementing the symmetry constraints of a 3-point function quite directly. On the contrary, the use of the four fundamental solutions introduced in \eqref{sol} and \eqref{solutions} requires significant manipulations (see the discussion in \cite{Coriano:2013jba}) in order to obtain the same result. In the case of 4-point functions, the possibility of expressing the dual conformal/conformal solution in the form of a 4K integral appears to be the natural generalization of such previous approach, and it also allows to discuss the kinematical limits in which the dynamics of a 4-point function reduces to that of a 3-point one. \\  
One can show that  hypergeometric functions of 3-variables, which belong to the class of Lauricella functions, can be related to 4K integrals. \\
 The key identity necessary to obtain the relation between the Lauricella functions and the 4K integral is given by the expression
	\begin{align}
\int_0^\infty dx\,x^{\a-1}\prod_{j=1}^3\,J_{\m_j}(a_j\,x)\,K_{\nu}(c\,x)&=2^{\a-2}\,c^{-\a-\l}\,\Gamma\left(\frac{\a+\l-\n}{2}\right)\Gamma\left(\frac{\a+\l+\n}{2}\right)\notag\\
&\hspace{-3cm}\times\prod_{j=1}^3\,\frac{a_j^{\m_j}}{\Gamma(\m_j+1)}F_C\left(\frac{\a+\l-\n}{2},\frac{\a+\l+\n}{2},\m_1+1,\m_2+1,\m_3+1;-\frac{a_1^2}{c^2},-\frac{a_2^2}{c^2},-\frac{a_3^2}{c^2}\right)\notag\\
&\centering\,\bigg[\l=\sum_{j=1}^3\,\m_j\,;\, \Re(\a+\l)>|\Re(\n)|,\,\Re(c)>\sum_{j=1}^{3}|\Im\,a_j|\bigg]\label{Prudnikov}.
\end{align}
If we rewrite the solutions of such systems in the form
	\begin{align}
	I_{\a-1\{\n_1,\n_2,\n_3,\n_4\}}(a_1,a_2,a_3,a_4)&=\int_0^\infty\,dx\,x^{\a-1}\,\prod_{i=1}^4(a_i)^{\n_i}\,K_{\n_i}(a_i\,x)
	\label{4Kintegral}
	\end{align} 
	with the Bessel functions $I_\nu,J_\nu, K_\nu$ related by the identities
	\begin{align}
	I_\nu(x)&=i^{-\n}\,J_{\n}(i\,x)\\
	K_\nu(x)&=\frac{\pi}{2\sin(\pi\,\n)}\bigg[I_{-\n}(x)-I_\n(x)\bigg]=\frac{1}{2}\bigg[i^\nu\, \Gamma(\n)\Gamma(1-\n)\,J_{-\n}(i\,x)+i^{-\n}\,\Gamma(-\n)\Gamma(1+\n)\,J_\n(i\,x)\bigg]\label{Kscomp}
	\end{align}
	where we have used the properties of the Gamma functions
	\begin{equation}
	\frac{\pi}{\sin(\pi\n)}=\Gamma(\n)\,\Gamma(1-\n),\qquad-\frac{\pi}{\sin(\pi\n)}=\Gamma(-\n)\,\Gamma(1+\n),	\end{equation}
	the dilatation Ward identities in this case can be written as
	\begin{equation}
	\bigg[(\D_t-3d)-\sum_{i=1}^4p_i\frac{\partial}{\partial p_i}\bigg]I_{\a\{\b_1,\b_2,\b_3,\b_4\}}(p_1,p_2,p_3,p_4)=0.
	\end{equation}
Using some properties of 4K integrals \cite{Maglio:2019grh} one can derive the relation
	\begin{align}
	(\a-\b_t+1+\D_t-3d)I_{\a\{\b_1,\b_2,\b_3,\b_4\}}(p_1,p_2,p_3,p_4)=0
	\end{align}
	where $\Delta_t=\sum_{i=1}^4 \Delta_i$,
	which is identically satisfied if the $\a$ exponent is equal to $\tilde{\a}$
	\begin{equation}
	\tilde{\a}=\b_t+3d-\D_t-1.
	\end{equation}
	The conformal Ward identities can then be re-expressed in the form
	\begin{equation}
	\left\{
	\begin{aligned}
	\textup{K}_{14}I_{\tilde{\a}\{\b_1,\b_2,\b_3,\b_4\}}&=0\\
	\textup{K}_{24}I_{\tilde{\a}\{\b_1,\b_2,\b_3,\b_4\}}&=0\\
	\textup{K}_{34}I_{\tilde{\a}\{\b_1,\b_2,\b_3,\b_4\}}&=0,
	\end{aligned}
	\right.
	\end{equation}
	generating the final relations
	\begin{equation}
	\left\{
	\begin{aligned}
	(d-2\D_4+2\b_4)I_{\tilde{\a}+1\{\b_1,\b_2,\b_3,\b_4-1\}}-(d-2\D_1+2\b_1)I_{\tilde{\a}+1\{\b_1-1,\b_2,\b_3,\b_4\}}&=0\\
	(d-2\D_4+2\b_4)I_{\tilde{\a}+1\{\b_1,\b_2,\b_3,\b_4-1\}}-(d-2\D_2+2\b_2)I_{\tilde{\a}+1\{\b_1,\b_2-1,\b_3,\b_4\}}&=0\\
	(d-2\D_4+2\b_4)I_{\tilde{\a}+1\{\b_1,\b_2,\b_3,\b_4-1\}}-(d-2\D_3+2\b_3)I_{\tilde{\a}+1\{\b_1,\b_2,\b_3-1,\b_4\}}&=0\\
	\end{aligned}
	\right.
	\end{equation}
	which are satisfied if
	\begin{align}
	\b_i=\D_i-\frac{d}{2},\qquad i=1,\dots,4
	\end{align}
	giving
	\begin{equation}
	\tilde{\a}=d-1.
	\end{equation}
	Finally, the final solution can be written as
	\begin{align}
	\phi(p_1,p_2,p_3,p_4)&=\bar{\bar{\a}}\, I_{d-1\left\{\D_1-\frac{d}{2},\D_2-\frac{d}{2},\D_3-\frac{d}{2},\D_4-\frac{d}{2}\right\}}(p_1,p_2,p_3,p_4)\notag\\
	&=\int_0^\infty\,dx\,x^{d-1}\,\prod_{i=1}^4(p_i)^{\D_i-\frac{d}{2}}\,K_{\D_i-\frac{d}{2}}(p_i\,x)
	,\label{4Kfin}
	\end{align}
	where $\bar{\bar{\a}}$ is a undetermined constant. \\
	Notice that given
	\begin{equation}
	I_{\a\{\b_1,\b_2,\b_3,\b_4\}}(p_1,p_2,p_3,p_4)=\int_0^\infty\,dx\,x^\a\,\prod_{i=1}^4(p_i)^{\b_i}\,K_{\b_i}(p_i\,x),
\end{equation}
its first derivative with respect the magnitudes of the momenta is given by
\begin{equation}
p_i\frac{\partial}{\partial p_i}I_{\a\{\b_j\}}=-p_i^2\,I_{\a+1\{\b_j-\d_{ij}\}},\qquad i,j=1,\dots,4.
\end{equation}
One can derive various relations satisfied by these types of integrals, as shown in \cite{Maglio:2019grh}. For instance, using
\begin{equation}
\int_0^\infty\,x^{\a+1}\frac{\partial}{\partial x}\left[\prod_{i=1}^4\,p_i^{\b_i}\,K_{\b_i}(p_i\,x)\right]=-\int_0^\infty\,\left[\frac{\partial x^{\a+1}}{\partial x}\right]\prod_{i=1}^4\,p_i^{\b_i}\,K_{\b_i}(p_i\,x)
\end{equation}
one derives the identity
\begin{equation}
\sum_{i=1}^{4}p_i^2I_{\a+1\{\b_j-\d_{ij}\}}=(\a-\b_t+1)\,I_{\a\{\b_j\}},\qquad j=1,\dots,4
\end{equation}
where $\b_t=\b_1+\b_2+\b_3+\b_4$. 
	One of the advantages of the use of the 4K integral expression of a solution is the simplified way by which the symmetry conditions can be imposed. In fact, by taking each of the 8 independent solutions identified in \eqref{oneeq}, and by rewriting them in the form of 4K integrals, the permutational symmetry of the correlators under the exchanges of the external momenta $p_i$ and scaling dimensions $\Delta_i$ becomes trivial.

	\section{Conclusions}
	We have reviewed recent progress on the analysis of the CWIs in momentum space for 3- and 4-point functions of ordinary CFTs in $d >2$. \\
	The momentum space approach, as mentioned in our introduction, allows to look at such theories from a very different perspective, which is simply not accessible from coordinate space. It allows to perform direct comparisons with the ordinary analysis of scattering amplitudes expressed in terms of Feynman diagrams, providing access to a wide class of methods which have been developed in this area of perturbative quantum field theories. \\
	The emergence of hypergeometric structures in the context of the CWI's is for sure an interesting feature of such equations which will be further explored in the near future, with interesting new results in this area. The study of the conformal phases quantum field theories is a fascinating topic which will probably receive continuing attention for its important physical applications and may shed light on several phenomena, ranging from  condensed matter theory to high energy theory. Therefore, our understanding of the fundamental mathematical structures which are part of these analysis and that can help in this process is of considerable importance.	
	
	 \vspace{1cm}

\centerline{\bf Acknowledgements} 
C.C. thanks the Institute for Theoretical Physics at ETH Zurich for hospitality while completing this work. We thank D. Theofilopoulos for collaborating on related studies. This work is partially supported by INFN under Iniziativa Specifica QFT-HEP.

\appendix
\section{Properties of triple-K integrals \label{appendixC}}

The modified Bessel function of the second kind is defined by
\begin{equation}
K_\n(x)=\frac{\pi}{2\sin(\n x)}[I_{-\n}(x)-I_{\n}(x)],\ \ \n\in\mathbb Z.
\end{equation}
If $\n=\frac{1}{2}+n$, for $n$ integer, the Bessel function can be expressed in the form
\begin{equation}
K_\n(x)=\sqrt{\frac{\pi}{2}}\,\frac{e^{-x}}{\sqrt{x}}\,\sum_{j=0}^{\lfloor\,|\n|-1/2\rfloor}\ \frac{(|\n|-1/2+j)!}{j!(|\n|-1/2-j)!}\frac{1}{(2x)^j},\ \ \n+1/2\in\mathbb{Z},
\end{equation}
where we have used the floor function ($\lfloor\, ,\rfloor$).  In particular
\begin{eqnarray}
&\hspace{-2cm}K_{\frac{1}{2}}(x)=\sqrt{\frac{\pi}{2}}\frac{e^{-x}}{\sqrt{x}},\quad &K_{\frac{3}{2}}(x)=\sqrt{\frac{\pi}{2}}\frac{e^{-x}}{\sqrt{x^3}}(1+x),\notag\\
&K_{\frac{5}{2}}(x)=\sqrt{\frac{\pi}{2}}\frac{e^{-x}}{\sqrt{x^5}},(x^2+3x+3),\quad
&K_{\frac{7}{2}}(x)=\sqrt{\frac{\pi}{2}}\frac{e^{-x}}{\sqrt{x^5}}(x^3+6x^3+15x+5).
\end{eqnarray}
Using this expressions the triple-K integrals can be calculated in a very simple way. For example, considering the case of half-integers $\b_i$, the 3K integral takes the form
\begin{align}
I_{\a\{\b_1\,\b_2,\b_3\}}&=\int_0^\infty\,dx\,x^\a\,p_1^{\b_1}\,p_2^{\b_2}\,p_3^{\b_3}\,K_{\b_1}(p_1x)\,K_{\b_2}(p_2x)\,K_{\b_3}(p_3x)\notag\\[1.5ex]
&\hspace{-1cm}=\sum_{k_1=0}^{|\b_1|-\frac{1}{2}}\ \,\sum_{k_2=0}^{|\b_2|-\frac{1}{2}}\ \,\sum_{k_3=0}^{|\b_3|-\frac{1}{2}}\ p_1^{\b_1-\frac{1}{2}-k_1}\,p_2^{\b_2-\frac{1}{2}-k_2}\,p_3^{\b_3-\frac{1}{2}-k_3}\,p_t^{k_t-\a-\frac{1}{2}}\,C_{k_1}(\b_1)\,C_{k_2}(\b_2)\,C_{k_3}(\b_3)\,\Gamma\left(\a-k_t-\frac{1}{2}\right),\label{halfinteg}
\end{align}
where $k_t=k_1+k_2+k_3$ and $p_t=p_1+p_2+p_3$ and we have define $C_{k_i}(\b_i)$ as
\begin{equation}
C_{k_i}(\b_i)\equiv \sqrt{\frac{\pi}{2^{2k_i+1}}}\,\frac{\left(|\b_i|-1/2+k_i\right)\,!}{k_i\,!\,\left(|\b_i|-1/2-k_i\right)\,!},
\end{equation}
and we have used the definition of the gamma function in order to write the integral
\begin{equation}
\int_0^\infty\,dx\,x^{\a-k_t-\frac{3}{2}}\,e^{-p_t\,x}=p_t^{k_t-\a+\frac{1}{2}}\int_0^\infty\,dy\,y^{\a-k_t-\frac{3}{2}}\,e^{-\,y}=p_t^{k_t-\a+\frac{1}{2}}\,\Gamma\left(\a-k_t-\frac{1}{2}\right).
\end{equation}
Using \eqref{halfinteg} we can calculate for instance the integrals
\begin{align}
I_{\frac{9}{2}\left\{\frac{3}{2},\frac{3}{2},-\frac{1}{2}\right\}}&=\left(\frac{\pi}{2}\right)^{3/2}\frac{3(p_1^2+p_2^2)+p_3^2+12p_1\,p_2+4p_3(p_1+p_2)}{p_3(p_1+p_2+p_3)^4}\\[1.2ex]
I_{\frac{7}{2}\left\{\frac{3}{2},\frac{3}{2},\frac{1}{2}\right\}}&=\left(\frac{\pi}{2}\right)^{3/2}\frac{2(p_1^2+p_2^2)+p_3^2+6 p_1\,p_2+3p_3(p_1+p_2)}{p_3(p_1+p_2+p_3)^3}
\end{align}
and any integrals with half-integer $\b_j$, with $j=1,2,3$.


\end{document}